%% file: res-scale-vpkiaas.tex
\documentclass[sigconf]{acmart}

\def\BibTeX{{\rm B\kern-.05em{\sc i\kern-.025em b}\kern-.08emT\kern-.1667em\lower.7ex\hbox{E}\kern-.125emX}}    

\usepackage{lipsum}

\usepackage{xcolor,colortbl}

\definecolor{Gray}{gray}{0.85}
\definecolor{LightCyan}{rgb}{0.88,1,1}
\newcolumntype{a}{>{\columncolor{Gray}}c}
\newcolumntype{b}{>{\columncolor{white}}c}

\usepackage[caption=false]{subfig}

\usepackage[nolist]{acronym}
\input{./acronym.tex}

\usepackage{algpseudocode}

\usepackage{setspace}

\errorcontextlines\maxdimen

\makeatletter
\newcommand*{\algrule}[1][\algorithmicindent]{\makebox[#1][l]{\hspace*{.5em}\thealgruleextra\vrule height \thealgruleheight depth \thealgruledepth}}%
\newcommand*{\thealgruleextra}{}
\newcommand*{\thealgruleheight}{.75\baselineskip}
\newcommand*{\thealgruledepth}{.25\baselineskip}

\newcount\ALG@printindent@tempcnta
\def\ALG@printindent{%
	\ifnum \theALG@nested>0
	\ifx\ALG@text\ALG@x@notext
	\else
	\unskip
	\addvspace{-1pt}
	\ALG@printindent@tempcnta=1
	\loop
	\algrule[\csname ALG@ind@\the\ALG@printindent@tempcnta\endcsname]%
	\advance \ALG@printindent@tempcnta 1
	\ifnum \ALG@printindent@tempcnta<\numexpr\theALG@nested+1\relax
	\repeat
	\fi
	\fi
}%
\usepackage{etoolbox}
\patchcmd{\ALG@doentity}{\noindent\hskip\ALG@tlm}{\ALG@printindent}{}{\errmessage{failed to patch}}
\makeatother

\newbox\statebox
\newcommand{\myState}[1]{%
	\setbox\statebox=\vbox{#1}%
	\edef\thealgruleheight{\dimexpr \the\ht\statebox+1pt\relax}%
	\edef\thealgruledepth{\dimexpr \the\dp\statebox+1pt\relax}%
	\ifdim\thealgruleheight<.75\baselineskip
	\def\thealgruleheight{\dimexpr .75\baselineskip+1pt\relax}%
	\fi
	\ifdim\thealgruledepth<.25\baselineskip
	\def\thealgruledepth{\dimexpr .25\baselineskip+1pt\relax}%
	\fi
	\State #1%
	\def\thealgruleheight{\dimexpr .75\baselineskip+1pt\relax}%
	\def\thealgruledepth{\dimexpr .25\baselineskip+1pt\relax}%
}

\usepackage{algorithm}

\makeatletter
\renewcommand{\ALG@beginalgorithmic}{\footnotesize} 
\makeatother

\usepackage{afterpage}
\newlength{\oldtextfloatsep}

\renewcommand\footnotetextcopyrightpermission[1]{} 
\begin{document}
\hyphenation{harmon-ization vulnera-ble invest-igated comm-uni-cations Con-sor-tium data-base ano-ny-mized ano-ny-mi-ty a-no-ny-mi-za-tion e-quip-ped pse-udo-ny-mi-ty co-lo-gne in-fra-struc-ture pseu-do-nyms pseu-do-nym informat-ion mecha-nism cryptogra-phic necessita-tes SEC-MACE non-over-lapping dy-na-mic-ally-sca-lable infra-struc-ture}

\title{Scaling Pseudonymous Authentication for Large Mobile Systems}

\author{Mohammad Khodaei}
\orcid{}
\affiliation{%
  \institution{Networked Systems Security Group}
  \streetaddress{KTH Royal Institute of Technology}
  \city{Stockholm} 
  \state{Sweden} 
}
\email{khodaei@kth.se}

\author{Hamid Noroozi}
\orcid{}
\affiliation{%
	\institution{Networked Systems Security Group}
	\streetaddress{KTH Royal Institute of Technology}
	\city{Stockholm} 
	\state{Sweden} 
}
\email{hnoroozi@kth.se}

\author{Panos Papadimitratos}
\affiliation{%
  \institution{Networked Systems Security Group}
  \streetaddress{KTH Royal Institute of Technology}
  \city{Stockholm} 
  \state{Sweden} 
}
\email{papadim@kth.se}


\begin{abstract}

The central building block of secure and privacy-preserving \ac{VC} systems is a \ac{VPKI}, which provides vehicles with multiple anonymized credentials, termed \emph{pseudonyms}. These pseudonyms are used to ensure message authenticity and integrity while preserving vehicle (thus passenger) privacy. In the light of emerging large-scale multi-domain \ac{VC} environments, the efficiency of the \ac{VPKI} and, more broadly, its scalability are paramount. By the same token, preventing misuse of the credentials, in particular, Sybil-based misbehavior, and managing \emph{``honest-but-curious''} insiders are other facets of a challenging problem. In this paper, we leverage a state-of-the-art \ac{VPKI} system and \emph{enhance} its functionality towards a highly-available, dynamically-scalable, and resilient design; this ensures that the system remains operational in the presence of benign failures or resource depletion attacks, and that it dynamically \emph{scales out}, or possibly \emph{scales in}, according to request arrival rates. Our full-blown implementation on the \acl{GCP} shows that deploying large-scale and efficient \ac{VPKI} can be cost-effective. 

\end{abstract}


\maketitle

\input{introduction}
\input{related-works}

\input{system-model-requirements}
\input{design}
\input{qualitative-analysis}

\input{quantitative-analysis}

\input{conclusions}
\input{acknowledgement}

\bibliographystyle{ACM-Reference-Format}
\bibliography{references} 

\input{appendix}

\end{document}

%% file: acronym.tex
\begin{acronym}
	\acro{AAA}{Authentication, Authorization and Accounting}
	\acro{ACL}{Access Control List}
	\acro{ACM}{AWS Certificate Manager}
	\acro{AES}{Advanced Encryption Standard}
	\acro{AKI}{Accountable Key Infrastructure}
	\acro{API}{Application Programming Interface}
	\acro{AWS}{Amazon Web Service}
	\acro{BSM}{Basic Safety Message}
	\acro{BYOD}{Bring Your Own Device}
	\acro{BF}{Bloom Filter}
	\acro{C2C-CC}{Car2Car Communication Consortium}
	\acro{C2I}{Car-to-Infrastructure}
	\acro{C$^2$RL}{Compressed \ac{CRL}}
	\acro{CA}{Certification Authority}
	\acro{CN}{Common Name}
	\acro{CAM}{Cooperative Awareness Message}
	\acro{CAMP VSC3}{Crash Avoidance Metrics Partnership Vehicle Safety Consortium}
	\acro{CDF}{Cumulative Distribution Function}
	\acro{CIA}{Confidentiality, Integrity and Availability}
	\acro{CRL}{Certificate Revocation List}
	\acro{CDN}{Content Delivery Network}
	\acro{COCA}{Cornell OnLine Certification Authority}
	\acro{CSR}{Certificate Signing Request}
	\acro{DAA}{Direct Anonymous Attestation}
	\acro{DDoS}{Distributed DoS}
	\acro{DDH}{Decisional Diffie-Helman}
	\acro{DENM}{Decentralized Environmental Notification Message}
	\acro{DHT}{Distributed Hash Table}
	\acro{DL/ECIES}{Discrete Logarithm and Elliptic Curve Integrated Encryption Scheme}
	\acro{DoT}{Department of Transportation}
	\acro{DoS}{Denial of Service}
	\acro{DoT}{Department of Transportation}
	\acro{DPA}{Data Protection Agency}
	\acro{DSRC}{Dedicated Short Range Communication}
	\acro{DSS}{Digital Signature Standard}
	\acro{DTLS}{Datagram \ac{TLS}}
	\acro{ECU}{Electronic Control Unit}
	\acro{EDR}{Event Data Recorder}
	\acro{ETSI}{European Telecommunications Standards Institute}
	\acro{ECDSA}{Elliptic Curve Digital Signature Algorithm}
	\acro{ECC}{Elliptic Curve Cryptography}
	\acro{EVITA}{E-safety Vehicle Intrusion protected Applications}
	\acro{FIPS}{Federal Information Processing Standard}
	\acro{FOT}{Field Operational Testing}
	\acro{FPGA}{Field-Programmable Gate Array}
	\acro{GCP}{Google Cloud Platform}
	\acro{GKE}{Google Kubernetes Engine}
	\acro{GPA}{Global Passive Adversary}
	\acro{GN}{GeoNetworking}
	\acro{GS-VLR}{Group Signatures with Verifier Local Revocation}
	\acro{GS}{Group Signatures}
	\acro{GM}{Group Manager}
	\acro{GBA}{Generic Bootstrapping Architecture}
	\acro{GNSS}{Global Navigation Satellite Systems}	
	\acro{GUI}{Graphic User Interface}
	\acro{HSM}{Hardware Security Module}
	\acro{HTTP}{Hypertext Transfer Protocol}
	\acro{HPA}{Horizontal Pod Autoscaler}
	\acro{IEEE}{Institute of Electrical and Electronics Engineers}
	\acro{IETF}{Internet Engineering Task Force}
	\acro{IoT}{Internet of Things}
	\acro{ITS}{Intelligent Transport System}
	\acro{IT}{Information Technologies}
	\acro{IMSI}{International Mobile Subscriber Identity}
	\acro{IMEI}{International Mobile Station Equipment Identity}
	\acro{IdP}{Identity Provider}
	\acro{IDS}{Intrusion Detection System}
	\acro{ISP}{Internet Service Provider}
	\acro{IAM}{Identity \& Access Management}
	\acro{KMS}{Key Management Service}
	\acro{LEA}{Law Enforcement Agency}
	\acro{LCPP}{Lightweight Conditional Privacy Preservation}
	\acro{LTC}{Long Term Certificate}
	\acro{LTCA}{Long Term CA}
	\acro{H-LTCA}{Home-LTCA}
	\acro{F-LTCA}{Foreign-LTCA}
	\acro{LDAP}{Lightweight Directory Access Protocol}
	\acro{LBS}{Location Based Service}
	\acro{LTE}{Long Term Evolution}
	\acro{LuST}{Luxembourg SUMO Traffic}
	\acro{MAC}{Message Authentication Code}
	\acro{MCA}{Message \ac{CA}}
	\acro{MEA}{Misbehavior Evaluation Authority}
	\acro{MANET}{Mobile Ad-hoc Network}
	\acro{MPB}{Most Pieces Broadcast}
	\acro{NoW}{Network on Wheel}
	\acro{OBU}{On-Board Unit}
	\acro{OEM}{Original Equipment Manufacturer}
	\acro{OCSP}{Online Certificate Status Protocol}
	\acro{PCA}{Pseudonym CA}
	\acro{PDP}{Policy Decision Point}
	\acro{PEP}{Policy Enforcement Point}
	\acro{QoS}{Quality of Service}
	\acro{PIR}{Private Information Retrieval}
	\acro{PKC}{Public Key Cryptography}
	\acro{PKCS}{Public Key Cryptosystem}
	\acro{PKI}{Public-Key Infrastructure}
	\acro{PRECIOSA}{Privacy Enabled Capability in Co-operative Systems and Safety Applications}
	\acro{PRESERVE}{Preparing Secure Vehicle-to-X Communication Systems}
	\acro{P2P}{peer-to-peer}
	\acro{PS}{Participatory Sensing}
	\acro{RA}{Resolution Authority}
	\acro{REST}{Representational State Transfer}
	\acro{RBAC}{Role Based Access Control}
	\acro{RCA}{Root \acs{CA}}
	\acro{RTC}{Real Time Clock}
	\acro{RDS}{Relational Database Service}
	\acro{RSU}{Roadside Unit}
	\acro{SAML}{Security Assertion Markup Language}
	\acro{SAS}{Sample Aggregation Service}
	\acro{SCMS}{Security Credential Management System}
	\acro{SCORE@F}{Système COopératif Routier Expérimental Français}
	\acro{SDSI}{Simple Distributed Security Infrastructure}
	\acro{SLA}{Service Level Agreement}
	\acro{SRAAC}{Secure Revocable Anonymous Authenticated Inter-Vehicle Communication}
	\acro{SeVeCom}{Secure Vehicle Communication}
	\acro{SIT}{Sichere Informationstechnologie}
	\acro{SLC}{Short-Lived Certificate}
	\acro{SoA}{Service-oriented-Approach}
	\acro{SIFS}{Short Inter Frame Space}
	\acro{SSO}{Single-Sign-On}
	\acro{SSL}{Secure Sockets Layer}
	\acro{SOAP}{Simple Object Access Protocol}
	\acro{TACK}{Temporary Anonymous Certified Key}
	\acro{TS}{Task Service}
	\acro{TLS}{Transport Layer Security}
	\acro{TPM}{Trusted Platform Module}
	\acro{TTP}{Trusted Third Party}
	\acro{TVR}{Ticket Validation Repository}
	\acro{URI}{Uniform Resource Identifier}
	\acro{VANET}{Vehicular Ad-hoc Network}
	\acro{V2I}{Vehicle-to-Infrastructure}
	\acro{V2V}{Vehicle-to-Vehicle}
	\acro{V2X}{\ac{V2V}/\ac{V2I}}
	\acro{VC}{Vehicular Communication}
	\acro{VM}{Virtual Machine}
	\acro{VSS}{\ac{VC} Security Subsystem}
	\acro{WAVE}{Wireless Access in Vehicular Environments}
	\acro{WSDL}{Web Services Discovery Language}
	\acro{W3C}{World Wide Web Consortium}
	\acro{V}{Vehicle}
	\acro{VANET}{Vehicular Ad-hoc Network}
	\acro{VLR}{Verifier-Local Revocation}
	\acro{VPKI}{Vehicular Public-Key Infrastructure}
	\acro{VPKIaaS}{VPKI as a Service}
	\acro{VM}{Virtual Machine}
	\acro{WS}{Web Service}
	\acro{WoT}{Web of Trust}
	\acro{WSACA}{\ac{WAVE} Service Advertisement \ac{CA}}
	\acro{XML}{Extensible Markup Language}
	\acro{XACML}{eXtensible Access Control Markup Language}
	\acro{3G}{3rd Generation}
\end{acronym}

%% file: introduction.tex
\section{Introduction}
\label{sec:scaling-vpkiaas-introduction}

\acresetall

In \ac{VC} systems, vehicles beacon \acp{CAM} and \acp{DENM} periodically, at high rates, to enable transportation safety and efficiency. It has been well-understood that \ac{VC} systems are vulnerable to attacks and that the privacy of their users is at stake. As a result, security and privacy solutions have been developed by standardization bodies (IEEE 1609.2 WG~\cite{1609-2016} and \acs{ETSI}~\cite{ETSI-102-638}), harmonization efforts (C2C-CC~\cite{c2c}), and projects (\acs{SeVeCom}~\cite{papadimitratos2008secure, SeVeCom, papadimitratos2007architecture}, \acs{PRESERVE}~\cite{preserve-url}, and CAMP~\cite{whyte2013security, US-VPKI}). A consensus towards using \ac{PKC} to protect \ac{V2X} communication is reached: a set of short-lived anonymized certificates, termed \emph{pseudonyms}, are issued by a \ac{VPKI}, e.g.,~\cite{whyte2013security, khodaei2014ScalableRobustVPKI, khodaei2018Secmace}, for registered vehicles. Vehicles switch from one pseudonym to a non-previously used one towards message unlinkability, as pseudonyms are per se inherently unlinkable. Pseudonymity is conditional, in the sense that the corresponding long-term vehicle identity can be retrieved by the \ac{VPKI} when needed, e.g., if vehicles deviating from system policies.

Deploying a \ac{VPKI} differs from a traditional \ac{PKI}, e.g.,~\cite{lets-encrypt, comodo, symantec}. One of the most important factors is the \ac{PKI} dimension, i.e., the number of registered ``users'' (vehicles) and the multiplicity of certificates per user. According to the US \ac{DoT}, a \ac{VPKI} should be able to issue pseudonyms for more that 350 million vehicles across the Nation~\cite{DOTHS812014}. Considering the average daily commute time to be 1 hour~\cite{DOTHS812014} and a pseudonym lifetime of 5 minutes, the \ac{VPKI} should be able to issue at least $1.5 \times 10^{12}$ pseudonyms per year, i.e., 5 orders of magnitude more than the number of credentials the largest current \ac{PKI} issues (10 million certificates per year~\cite{whyte2013security}). Note that this number could be even greater for the entire envisioned \acp{ITS} ecosystem, e.g., including pedestrians and cyclists, \acp{LBS}~\cite{ETSI-102-638, papadimitratos2009vehicular, shokri2014hiding} and vehicular social networks~\cite{jin2016security}. More so, outside the \ac{VC} realm, there is an ongoing trend towards leveraging short-lived certificates~\cite{topalovic2012towards} for the Internet: web servers request new short-lived certificates, valid for a few days~\cite{topalovic2012towards}. This essentially diminishes the vulnerability window, e.g., if a single \ac{CA} were compromised~\cite{topalovic2012towards}, or if a large fraction of certificates needed to be revoked after the latest \ac{CRL} was distributed among all entities~\cite{mcdaniel2000response, clark2013sok, khodaei2018VehicleCentric}.

With emerging large-scale multi-domain \ac{VC} environments~\cite{papadimitratos2009vehicular, khodaei2015VTMagazine, 1609-2016, ETSI-102-638, c2c}, the efficiency of the \ac{VPKI} and, more broadly, its scalability are paramount. Vehicles could request pseudonyms for a long period, e.g., 25 years~\cite{kumar2017binary}. However, extensive pre-loading with millions of pseudonyms per vehicle for a long period is computationally costly and inefficient in terms of utilization~\cite{khodaei2018Secmace}. Moreover, in case of revocation~\cite{mcdaniel2000response, clark2013sok, khodaei2018VehicleCentric}, a huge \ac{CRL} should be distributed among all vehicles due to long lifespan of the credentials, e.g.,~\cite{kumar2017binary}: a sizable portion of the \ac{CRL} is irrelevant to a receiving vehicle and can be left unused, i.e., wasting of significant bandwidth for \ac{CRL} distribution~\cite{khodaei2018VehicleCentric, simplicio2018acpc}. Alternatively, each vehicle could interact with the \ac{VPKI} regularly, e.g., once or a few times per day, not only to refill its pseudonym pool but also to fetch the latest revocation information\footnote{Note that Cellular-\ac{V2X} provides reliable and low-latency \ac{V2X} communication with a wide range of coverage~\cite{andrews2014will, agiwal2016next, abboud2016interworking}; thus, network connectivity will not be a bottleneck.}. However, the performance of a \ac{VPKI} system can be drastically degraded under a clogging \ac{DoS} attack~\cite{khodaei2014ScalableRobustVPKI, khodaei2018Secmace}, thus, compromising the availability of the \ac{VPKI} entities. Moreover, a \emph{flash crowd}~\cite{ari2003managing}, e.g., a surge in pseudonym acquisition requests during rush hours, could render the \ac{VPKI} unreachable, or drastically decrease its quality of service.

The cost of \ac{VPKI} unavailability is twofold: security (degradation of road safety) and privacy. An active malicious entity could prevent other vehicles from accessing the \ac{VPKI} to fetch the latest revocation information. Moreover, signing \acp{CAM} with the private keys corresponding to expired pseudonyms, or the \ac{LTC}, is insecure and detrimental to user privacy. Even though one can refill its pseudonym pool by relying on anonymous authentication primitives, e.g.,~\cite{calandriello2007efficient, PapadiCLH-C-08, calandriello2011performance, khodaei2017RHyTHM}, the performance of the safety-related applications could be degraded. For example, leveraging anonymous authentication schemes for the majority of vehicles results in causing 30\% increase in cryptographic processing overhead in order to validate \acp{CAM}~\cite{khodaei2017RHyTHM}. Thus, it is crucial to provide a highly-available, scalable, and resilient \ac{VPKI} design that could efficiently issue pseudonyms in an \emph{on-demand} fashion\footnote{Unlike issuing short-lived certificates~\cite{topalovic2012towards} for the Internet that responses can be cached, issuing on-demand pseudonyms cannot be precomputed: each vehicle requests new certificates with a different public key, important for unlinkability/privacy.}~\cite{ma2008pseudonym, khodaei2016evaluating}.

Considering a multi-domain development of \ac{VC} systems, with a multiplicity of service providers, each vehicle could obtain pseudonyms from various service providers. The acquisition of multiple simultaneously valid (sets of) pseudonyms would enable an adversary to inject multiple erroneous messages, e.g., hazard notifications, as if they were originated from multiple vehicles, or affect protocols based on voting, by sending out false, yet authenticated, information. Even though there are distributed schemes to identify Sybil~\cite{douceur2002sybil} nodes, e.g.,~\cite{xiao2006detection, golle2004detecting}, or mitigate this vulnerability by relying on \acp{HSM}~\cite{papadimitratos2007architecture}, a \ac{VPKI} system should prevent such credentials misuse on the infrastructure side, e.g.,~\cite{khodaei2014ScalableRobustVPKI, khodaei2018Secmace}. However, when deploying such a system, e.g.,~\cite{khodaei2018VPKIaaS, cincilla2016vehicular}, on the cloud, a malicious vehicle could repeatedly request pseudonyms; in fact, requests might be delivered to different replicas of a micro-service, releasing multiple simultaneously valid pseudonyms. Mandating a centralized database, shared among all replicas to ensure \emph{isolation} and \emph{consistency} of all transactions, would mitigate such a vulnerability. However, this contradicts highly efficient and timely pseudonyms provisioning for large-scale mobile systems.

\emph{Contributions:} In this paper, we leverage and \emph{enhance} a state-of-the-art \ac{VPKI}, and propose a \emph{\ac{VPKIaaS}} system towards a highly-available, dynamically-scalable, and fault-tolerant (highly-resilient) design, ensuring the system remains operational in the presence of benign failures or any resource depletion attack (clogging a \ac{DoS} attack). Moreover, our scheme eradicates Sybil-based misbehavior when deploying such a system on the cloud with multiple replicas of a micro-service without diminishing the pseudonym acquisition efficiency. All procedures of deployment and migration to the cloud, e.g., bootstrapping phase, initializing the micro-services, pseudonym acquisition process, monitoring health and load metrics, etc., are fully automated. Through extensive experimental evaluation, we show that the \ac{VPKIaaS} system could dynamically scale out, or possibly scale in\footnote{In the cloud terminology, scaling in/out, termed \emph{horizontal} scaling, refers to replicating a new instance of a service, while scaling up/down, termed \emph{vertical} scaling, refers to allocating/deallocating resources for an instance of a given service.}, based on the \ac{VPKIaaS} system workload and the requests' arrival rate, so that it can comfortably handle \emph{unexpected} demanding loads while being cost-effective by systematically allocating and deallocating resources. Our experimental evaluation shows a 36-fold improvement over prior work~\cite{cincilla2016vehicular}: the processing delay to issue 100 pseudonyms for~\cite{cincilla2016vehicular} is approx. 2010 ms, while it is approx. 56 ms in our system. Moreover, the performance of~\cite{khodaei2018Secmace} drastically decreases when there is a surge in the pseudonym request arrival rates; on the contrary, our \ac{VPKIaaS} system can comfortably handle demanding loads request while efficiently issuing batches of pseudonyms.

In the rest of the paper, we describe background and related work (Sec.~\ref{sec:scaling-vpkiaas-related-work}) and the system model and objectives (Sec.~\ref{sec:scaling-vpkiaas-model-requirement}). We then explain the \ac{VPKIaaS} system, detailing security protocols (Sec.~\ref{sec:scaling-vpkiaas-services-and-security-protocols}), and provide a qualitative analysis (Sec.~\ref{sec:scaling-vpkiaas-security-and-privacy-analysis}), followed by a quantitative analysis (Sec.~\ref{sec:scaling-vpkiaas-performance-evaluation}), before the conclusion (Sec~\ref{sec:scaling-vpkiaas-conclusion}).


%% file: related-works.tex
\section{Background and Related Work}
\label{sec:scaling-vpkiaas-related-work}

A \ac{VPKI} can provide vehicles with valid pseudonyms for a long period, e.g., 25 years~\cite{kumar2017binary}. However, extensive preloading with millions of pseudonyms per vehicle for such a long period is computationally costly, inefficient in terms of utilization and cumbersome for revocation~\cite{khodaei2015VTMagazine, khodaei2018VehicleCentric}. On the contrary, several proposals suggest more frequent Vehicle-to-\ac{VPKI} interactions, namely \emph{on-demand} schemes, e.g.,~\cite{fischer2006secure, schaub2010v, khodaei2014ScalableRobustVPKI, khodaei2018Secmace}. This strategy provides more efficient pseudonym utilization and revocation, thus being effective in fending off misbehavior. But, for on-demand pseudonym acquisition, one needs to design (and deploy) an efficient and scalable system while being resilient against any resource depletion attack. Even though \ac{VPKI} systems may handle large-scaled distributed scenarios, e.g.,~\cite{cincilla2016vehicular}, there is lack of dynamic scalability (i.e., dynamically scale out/in according to the arrival rates) and resilient to a resource depletion attack, e.g., a \ac{DDoS} attack. Beyond a significant performance improvement over~\cite{cincilla2016vehicular}, our \ac{VPKIaaS} implementation is highly-available, dynamically-scalable, and fault-tolerant.

Sybil-based~\cite{douceur2002sybil} misbehavior can seriously affect the operation of \ac{VC} systems, as multiple fabricated non-existing vehicles could pollute the network by injecting false information. For example, an adversary with multiple valid pseudonyms, termed here a \emph{Sybil} node, could create an illusion of traffic congestion towards affecting the operation of a traffic monitoring system, or broadcast fake misbehavior detection votes~\cite{raya2007eviction, ruj2011data, reidt2009fable}, or disseminate Spam to other users in a vehicular social network~\cite{jin2016security}. The idea of enforcing non-overlapping pseudonym lifetimes was first proposed in~\cite{papadimitratos2007architecture}. This prevents an adversary from equipping itself with multiple valid identities, and thus affecting protocols of collection of multiple inputs, e.g., based on voting, by sending out redundant false, yet authenticated, information. Even though this idea has been accepted, a number of proposals, e.g.,~\cite{whyte2013security, kumar2017binary}, do not prevent a vehicle from obtaining simultaneously valid pseudonyms via multiple pseudonym requests. The existence of multiple pseudonym issuers deteriorate the situation: a vehicle could request pseudonyms from multiple service providers, while each of them is not aware whether pseudonyms for the same period were issued by any other service provider. One can mitigate this vulnerability by relying on an \ac{HSM}, ensuring all signatures are generated under a single valid pseudonym at any time. There are also distributed schemes to detect Sybil nodes based on radio characteristics and triangulation, e.g.,~\cite{xiao2006detection, golle2004detecting}; such strategies are application-dependent, e.g., this cannot guarantee the operation of a traffic monitoring system from an adversary who disseminates multiple traffic congestion messages, each signed under a distinct ``fake'' private key.

V-tokens~\cite{schaub2010v} prevents a vehicle from obtaining multiple simultaneously valid pseudonyms due to having service providers communicating with each other, e.g., a distributed hash table. SECMACE~\cite{khodaei2018Secmace} (including its predecessors~\cite{khodaei2014ScalableRobustVPKI, khodaei2016evaluating}) prevents Sybil-based misbehavior on the infrastructure side without the need for an additional entity, i.e., extra interactions or intra-\ac{VPKI} communications. More specifically, it ensures each vehicle has one valid pseudonym at any time in a multi-domain environment. However, when deploying such a system on the cloud, a malicious vehicle could repeatedly request pseudonyms, hoping that requests are delivered to different replicas of a micro-service, thus obtaining multiple simultaneously valid pseudonyms, e.g.,~\cite{khodaei2018VPKIaaS, cincilla2016vehicular}. Unlike such schemes, our \ac{VPKIaaS} scheme prevents Sybil-based misbehavior on the cloud-deployed infrastructure: it ensures that each vehicle can only have one valid pseudonym at any time in a multi-domain \ac{VC} environment; more important, it does not affect timely issuance of pseudonyms.

The \ac{VPKI} entities are, often implicitly, assumed to be fully trustworthy. Given the experience from recent mobile applications, e.g., \cite{comodo-hacked, diginotar-hacked, google-cert-hacked}, the adversarial model is extended from fully trustworthy to \emph{honest-but-curious} \ac{VPKI} servers, notably in~\cite{khodaei2018Secmace, whyte2013security}. Such honest-but-curious entities may subvert the security protocols and deviate from system policies if gained an advantage without being identified, e.g., inferring user sensitive information~\cite{wiedersheim2010privacy, khodaei2018PrivacyUniformity, khodaei2018poster, vaas2018nowhere}. Outside the \ac{VC} realm, there are different proposals for \ac{PKI} to be resilient against \emph{compromised} insiders. Such schemes rely on signing a certificate by more than a threshold number of \acsp{CA}, e.g.,~\cite{kim2013accountable, dykcik2018blockpki}; however, such schemes cannot be used by \ac{VC} systems. For example, issuing a certificate in~\cite{dykcik2018blockpki} takes approximately 2~minutes and it varies with the number of required \acsp{CA}. Obviously, this contradicts with \emph{on-demand} pseudonym acquisition strategies for \ac{VC} systems, e.g.,~\cite{ma2008pseudonym, khodaei2016evaluating, khodaei2017RHyTHM, khodaei2018Secmace}, which necessitate efficient pseudonym provisioning.


%% file: system-model-requirements.tex
\section{System Model and Objectives}
\label{sec:scaling-vpkiaas-model-requirement}

\subsection{Overview and Assumptions}
\label{subsec:scaling-vpkiaas-overview-assumptions}

A \ac{VPKI} consists of a set of Certification Authorities (CAs) with distinct roles: the \ac{RCA}, the highest-level authority, certifies other lower-level authorities; the \ac{LTCA} is responsible for the vehicle registration and the \acf{LTC} issuance, and the \ac{PCA} issues pseudonyms for the registered vehicles. Pseudonyms have a lifetime (a validity period), typically ranging from minutes to hours; in principle, the shorter the pseudonym lifetime is, the higher the unlinkability and thus the higher privacy protection can be achieved. We assume that each vehicle is registered only with its \emph{\ac{H-LTCA}}, the \emph{policy decision and enforcement point}, reachable by the registered vehicles. Without loss of generality, a \emph{domain} can be defined as a set of vehicles in a region, registered with the \ac{H-LTCA}, subject to the same administrative regulations and policies~\cite{papadimitratos2006securing, khodaei2015VTMagazine}. There can be several \acp{PCA}, each active in one or more domains; any legitimate, i.e., registered, vehicle is able to obtain pseudonyms from any \ac{PCA}, the pseudonym provider (as long as there is a trust established between the two \acsp{CA}). Trust between two domains can be established with the help of the \ac{RCA}, or through cross certification.

\begin{figure}[!t] 
	\vspace{-0em}
	\centering
	\includegraphics[width=0.47\textwidth,height=0.47\textheight,keepaspectratio] {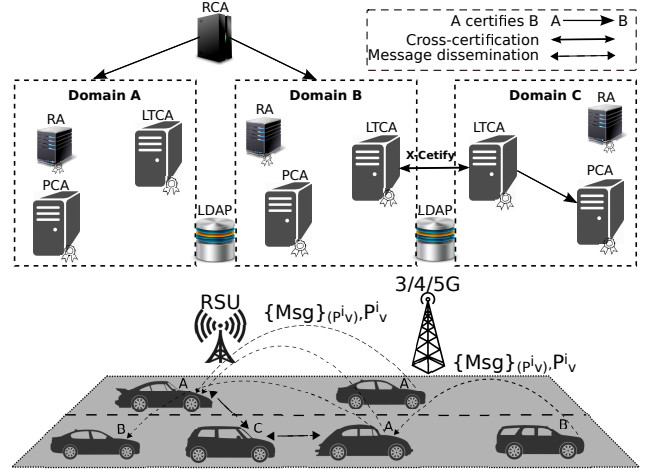} 
	\vspace{-0.75em}
	\caption{A \ac{VPKI} Overview for Multi-domain \ac{VC} Systems.}
	\label{fig:scaling-vpkiaas-system-overview}
	\vspace{-1.5em}
\end{figure}

Each vehicle interacts with the \ac{VPKI} entities to obtain a batch of pseudonyms, each having a corresponding short-term private key, to sign and disseminate their mobility information, e.g., \acp{CAM} or \acp{DENM}, time- and geo-stamped, periodically or when needed as a response to a specific event. Fig.~\ref{fig:scaling-vpkiaas-system-overview} shows an overview of a \ac{VPKI} with three domains, $A$, $B$ and $C$. Domains $A$ and $B$ have established trust with the help of a higher level authority, i.e., the \ac{RCA}, while domains $B$ and $C$ have established security association by cross certification. The vehicles in the figure are labeled with the domains they are affiliated to. A vehicle registered in domain $A$ digitally signs outgoing messages with the private key, $k^i_v$, corresponding to $P^i_v$, which signifies the current valid pseudonym signed by the \ac{PCA}. The pseudonym is then attached to the signed messages to enable verification by any recipient. Upon reception, the pseudonym is verified before the message itself (signature validation). This process ensures communication authenticity, message integrity, and non-repudiation. Vehicles switch from one pseudonym to another one (non-previously used) to achieve unlinkability, thus protecting sender's privacy as the pseudonyms are inherently unlinkable.

Each vehicle \emph{``decides''} when to trigger the pseudonym acquisition process based on various factors~\cite{khodaei2016evaluating}. Such a scheme requires sparse connectivity to the \ac{VPKI}, but it facilitates an \ac{OBU} to be \emph{preloaded} with pseudonyms proactively, covering a longer period, e.g., a week or a month, should the connectivity be expected heavily intermittent. A universally fixed interval, $\Gamma$, is specified by the \ac{H-LTCA} and all pseudonyms in that domain are issued with the lifetime ($\tau_{P}$) aligned with the \ac{VPKI} clock~\cite{khodaei2018Secmace}. As a result of this policy, at any point in time, all the vehicles transmit using pseudonyms that cannot be distinguished based on their issuance time thanks to this time alignment.

All vehicles (\acp{OBU}) registered are equipped with \acp{HSM}, ensuring that private keys never leave the \ac{HSM}. Moreover, we assume that there is a misbehavior detection system, e.g.,~\cite{bissmeyer2014misbehavior}, that triggers revocation. The \acf{RA} can initiate a process to resolve and revoke all pseudonyms of a misbehaving vehicle~\cite{Papadi:C:08}: it interacts with the corresponding \acp{PCA} and \ac{LTCA} (a detailed protocol description, e.g., in~\cite{khodaei2014ScalableRobustVPKI, khodaei2018Secmace}) to resolve and revoke all credentials issued for a misbehaving vehicle. Consequently, the misbehaving vehicle can no longer obtain credentials from the \ac{VPKI}. The \ac{VPKI} is responsible for distributing the \acp{CRL} and notifying all legitimate entities about the revocation, e.g.,~\cite{khodaei2018VehicleCentric}. We further assume that the cloud service providers are honest and they provide a service with the desired \ac{SLA}; in terms of secret management, we assume that the cloud service providers are fully trustworthy.

\subsection{Adversarial Model and Requirements}
\label{subsec:scaling-vpkiaas-adversarial-model}

We extend the general adversary model in secure vehicular communications~\cite{papadimitratos2006securing, khodaei2018Secmace} to include an \emph{honest-but-curious} service provider, i.e., a \ac{PCA} that attempts to gain advantages towards its goal, e.g., profiling users. In addition, in the context of this work, malicious \acp{PCA} could try to (i) issue multiple sets of (simultaneously valid) pseudonyms for a legitimate vehicle, or (ii) issue a set of pseudonyms for a non-existing (illegitimate) vehicle, or (iii) fraudulently accuse different vehicles (users) during a pseudonym resolution process. A deviant \ac{LTCA} could attempt to map a different \ac{LTC} during the resolution process, thus misleading the system. In our adversarial model, we assume that the \ac{LTCA} does not misbehave by unlawfully registering illegitimate vehicles, i.e., issuing fake \acp{LTC}, but it could be tempted to issue fake \emph{authorization tickets}, to be used during pseudonym acquisition process\footnote{During the registration process, the \ac{H-LTCA} registers a vehicle upon receiving a request from the corresponding \ac{OEM}, i.e., to fraudulently register a vehicle, two entities must collude. But, in order to issue a fake ticket, the \ac{H-LTCA} could do it without interacting with any other entity.}. The \ac{RA} can also continuously initiate pseudonym validation process towards inferring user sensitive information. Our adversarial model considers multiple \ac{VPKI} servers collude, i.e., share information that each of them individually infers with the others, to harm user privacy.

In a multi-\ac{PCA} environment, malicious (compromised) clients raise two challenges. First, they could repeatedly request multiple simultaneously valid pseudonyms, thus misbehaving each as multiple registered legitimate-looking vehicles. Second, they could degrade the operations of the system by mounting a clogging \ac{DoS} attack against the \ac{VPKI} servers. \emph{External adversaries}, i.e., unauthorized entities, could try to harm the system operations by launching a \ac{DoS} (or a \ac{DDoS}) attack, thus degrading the availability of the system. But they are unable to successfully forge messages or `crack' the employed cryptosystems and cryptographic primitives.

Security and privacy requirements for \ac{V2X} communications have been extensively specified in~\cite{papadimitratos2006securing}, and additional requirements for \ac{VPKI} entities in~\cite{khodaei2018Secmace} and the \ac{CRL} distribution in~\cite{khodaei2018VehicleCentric}. Beyond the aforementioned requirements, we need to thwart Sybil-based attacks when deploying \ac{VPKIaaS} system on the cloud (without degrading efficient pseudonym issuance). At the same time, we need to ensure that the \ac{VPKIaaS} system is highly-available and dynamically-scalable: the system \emph{dynamically} scales out, or possibly scales in, according to the requests' arrival rate, to handle any demanding load while being cost-effective by systematically allocating and deallocating resources. Moreover, we need to ensure that the scheme is resilient to any resource depletion attack.


%% file: design.tex
\section{\ac{VPKI} Services Overview \& Security Protocols}
\label{sec:scaling-vpkiaas-services-and-security-protocols}

In the registration phase, each \ac{H-LTCA} registers vehicles within its domain and maintains their long-term identities. At the bootstrapping phase, each vehicle needs to discover the \ac{VPKI}-related information, e.g., the available \acp{PCA} in its home domain, or the desired \ac{F-LTCA} and \acp{PCA} in a foreign domain, along with their corresponding certificates. To facilitate the overall intra-domain and multi-domain operations, a vehicle first finds such information from a \ac{LDAP}~\cite{sermersheim2006lightweight} server. This is carried out without disclosing the real identity of the vehicle. We presume connectivity to the \ac{VPKI}, e.g., via \acp{RSU} or Cellular-\ac{V2X}; should the connectivity be intermittent, vehicle, i.e., the \ac{OBU}, could initiate pseudonym provisioning proactively based on different parameters, e.g., the number of remaining valid pseudonyms, the residual trip duration, and the networking connectivity.

The \ac{H-LTCA} authenticates and authorizes vehicles over a mutually authenticated \ac{TLS}~\cite{dierks2008transport} tunnel. This way the vehicle obtains a \emph{native ticket} ($n\textnormal{-}tkt$) from its \ac{H-LTCA} while the targeted \ac{PCA} or the actual pseudonym acquisition period is hidden from the \ac{H-LTCA}; the ticket is anonymized and it does not reveal its owner's identity (Protocol~\ref{protocol:scaling-vpkiaas-requesting-ticket-algorithm} and Protocol~\ref{protocol:scaling-vpkiaas-issuing-ticket-algorithm} in the Appendix). The ticket is then presented to the intended \ac{PCA}, over a unidirectional (server-only) authenticated \ac{TLS}, to obtain pseudonyms (Protocol~\ref{protocol:scaling-vpkiaas-issuing-psnyms}).

When the vehicle travels in a foreign domain, it should obtain new pseudonyms from a \ac{PCA} operating in that domain; otherwise, the vehicle would stand out and be more easily traceable (linkable). The vehicle first requests a \emph{foreign ticket} ($f\textnormal{-}tkt$) from its \ac{H-LTCA} (without revealing its targeted \ac{F-LTCA}) so that the vehicle can be authenticated and authorized by the \ac{F-LTCA}. In turn, the \ac{F-LTCA} provides the vehicle with a new ticket ($n\textnormal{-}tkt$), which is native within the domain of the \ac{F-LTCA} to be used for pseudonym acquisition in that (foreign) domain. The vehicle then interacts with its desired \ac{PCA} to obtain pseudonyms. Obtaining an $f\textnormal{-}tkt$ is transparent to the \ac{H-LTCA}: the \ac{H-LTCA} cannot distinguish between native and foreign ticket requests. This way, the \ac{PCA} in the foreign domain cannot distinguish native requesters from foreign ones. For liability attribution, our scheme enables the \ac{RA}, with the help of the \ac{PCA} and the \ac{LTCA}, to initiate a resolution process, i.e., to resolve a pseudonym to its long-term identity. Each vehicle can interact with any \ac{PCA}, within its home or a foreign domain, to fetch the \ac{CRL}~\cite{khodaei2018VehicleCentric} and perform \ac{OCSP}~\cite{khodaei2014ScalableRobustVPKI} operations, authenticated with a current valid pseudonym.

\subsection{VPKI\lowercase{ as a }Service (VPKI\lowercase{aa}S)}
\label{subsec:scaling-vpkiaas-vpkiaas}

\begin{figure} [!t]
	\vspace{-2.5em}
	\begin{center}
		\centering
		\includegraphics[trim=0cm 0cm 0cm 0cm, clip=true, totalheight=0.24\textheight,angle=0]{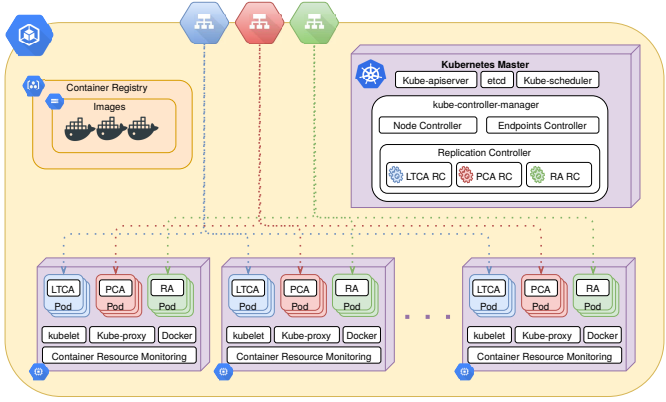}
		\vspace{-1.5em}
		\caption{A High-level Overview of \acs{VPKIaaS} Architecture.}
		\label{fig:scaling-vpkiaas-system-design}
	\end{center}
\end{figure}

We migrate the \ac{VPKI} on the \ac{GCP}~\cite{gcp} for the availability, reliability, and dynamic scalability of the \ac{VPKI} system under various circumstances. Fig.~\ref{fig:scaling-vpkiaas-system-design} illustrates a high-level abstraction of the \acs{VPKIaaS} architecture on a managed Kubernetes cluster~\cite{kubernetes} on \ac{GCP}.\footnote{Note that the \ac{RCA} entity is assumed to be off-line, thus not included in this abstraction.} A set of Pods will be created for each micro-service, e.g., \ac{LTCA} or \ac{PCA}, from their corresponding container images, facilitating their horizontal scalability. When the rate of pseudonym requests increases, the Kubernetes master, shown on the top in Fig.~\ref{fig:scaling-vpkiaas-system-design}, schedules new Pods or kills a running Pod in case of benign failures, e.g., system faults or crashes, or resource depletion attacks, e.g., a \ac{DoS} attack. The Pods could be scaled out to the number, set in the deployment configuration, or scaled out to the amount of available resources enabled by Kubernetes nodes.

Each Pod publishes two types of metrics: \emph{load} and \emph{health}. The load metric values are generated by a resource monitoring service, which facilitates horizontal scaling of a micro-service: upon reaching a threshold of a pre-defined load, replication controller replicates a new instance of the micro-service to ensure a desired \ac{SLA}. The health metric ensures correct operation of a micro-service by persistently monitoring its status: a faulty Pod is killed and a new one is created. In our \ac{VPKIaaS} system, we define CPU usages as the load metric. In order to monitor the health condition of a micro-service, dummy requests (dummy tickets for the \ac{LTCA} micro-services and dummy pseudonyms for the \ac{PCA} micro-services) are locally queried by each Pod\footnote{A dummy ticket request is constructed by an \ac{LTCA} Pod to validate the correctness of ticket issuance procedure while a dummy pseudonym request is constructed by a \ac{PCA} Pod to ensure the correctness of pseudonym issuance procedure. Such dummy requests cannot be used by a compromised Pod to issue fake pseudonyms (see Sec.~\ref{sec:scaling-vpkiaas-security-and-privacy-analysis}).}.

\subsection{Security Protocols}
\label{subsec:scaling-vpkiaas-security-protocols}

In this section, we provide the detailed description of pseudonym acquisition processes (Protocol~\ref{protocol:scaling-vpkiaas-issuing-psnyms}) and pseudonym issuance validation process (Protocol~\ref{protocol:scaling-vpkiaas-plausibility-check-for-the-pseudonym-issuance}) in order to identify misbehaving \ac{PCA} issuing fraudulent pseudonym. Furthermore, in order to mitigate Sybil attacks on the side of \ac{VPKIaaS} system, we propose two protocols (Protocols~\ref{protocol:scaling-vpkiaas-ticket-request-validation-by-redis} and~\ref{protocol:scaling-vpkiaas-pseudonym-request-validation-by-redis}): an in-memory key-value Redis database~\cite{redis} is shared among all replicas of a micro-service, to facilitate efficient validation of tickets and pseudonyms requests. Table~\ref{table:scaling-vpkiaas-protocols-notation} shows the notation used in the security protocols.

\subsubsection{\textbf{Pseudonym Acquisition Process (Protocol~\ref{protocol:scaling-vpkiaas-issuing-psnyms})}}
\label{subsubsec:dos-res-vpkiaas-pseudonym-acquisition-process}

Each vehicle first requests an anonymous ticket~\cite{khodaei2016evaluating, khodaei2014ScalableRobustVPKI} from its \ac{H-LTCA}, using it to interact with the desired \ac{PCA} to obtain pseudonyms; due to lack of space, we provide the detailed ticket acquisition process in Appendix. Upon reception of a valid ticket, it generates \acp{CSR} with \ac{ECDSA} public/private key pairs~\cite{1609-2016, ETSI-102-638} and sends the request to the \ac{PCA}. Vehicle-\ac{LTCA} is over mutually authenticated \ac{TLS}~\cite{dierks2008transport} tunnels (or \ac{DTLS}~\cite{rescorla2012datagram}) while the vehicle-\ac{PCA} communication is over a unidirectional (server-only) authenticated \ac{TLS} (or \acs{DTLS}); this ensures that the \ac{PCA} does not infer the actual identity of the requester.

\begin{table} [!t]
	\vspace{-2.5em}
	\caption{Notation used in the protocols}
	\vspace{-1em}
	\centering
	\resizebox{0.49\textwidth}{!}
	{
		\renewcommand{\arraystretch}{1.1}
		\begin{tabular}{l | *{1}{c} r}
			\hline \hline
			$(P^{i}_{v})_{pca}$, $P^{i}_{v}$ & a pseudonym signed by the \acs{PCA} \\\hline
			$(LK_v, Lk_v)$ & long-term public/private key pairs \\\hline
			$(K^i_v, k^i_v)$ & \shortstack{pseudonymous public/private key pairs} \\\hline 
			$Id_{req}, Id_{res}, Id_{ca}$ & request/response/CA unique identifiers \\ \hline
			$(msg)_{\sigma_{v}}$ & a signed message with the vehicle's private key \\\hline
			$N, Rnd$ & nonce, a random number \\\hline
			$t_{now}, t_s, t_e$ & fresh/current, starting, and ending timestamps \\\hline
			$n\textnormal{-}tkt$, $f\textnormal{-}tkt$ & native ticket, foreign ticket \\\hline 
			$H()$ & hash function \\\hline
			$Sign(Lk_{}, msg)$ & signing a message with the private key ($Lk$) \\\hline 
			$Verify(LK_{}, msg)$ & verifying a message with the public key \\ \hline 
			$\tau_{P}$ & pseudonym lifetime \\\hline
			$\Gamma$ & interacting interval with the \ac{VPKI} \\\hline 
			$IK$ & identifiable key \\\hline
			$V$ & vehicle \\\hline
			$\zeta, \chi$ & temporary variables \\\hline 
			\hline
		\end{tabular}
		\renewcommand{\arraystretch}{1}
		\label{table:scaling-vpkiaas-protocols-notation}
	}
	\vspace{-1em}
\end{table}

\setlength{\textfloatsep}{0pt}
\begin{algorithm}[t!]
	\floatname{algorithm}{Protocol}
	\caption{Issuing Pseudonyms (by the \acs{PCA})}
	\label{protocol:scaling-vpkiaas-issuing-psnyms}
	\algloop{For}{}
	
	\algblock{Begin}{End}
	\begin{algorithmic}[1]
		\Procedure{IssuePsnyms}{$Req$}
		\myState {\scriptsize $Req \hspace{-0.35em} \to \hspace{-0.35em} {(Id_{req}, \hspace{-0.1em} Rnd_{n\textnormal{-}tkt}, \hspace{-0.1em} tkt_{\sigma_{ltca}}, \hspace{-0.1em} \{(K^1_v)_{\sigma_{k^1_v}}, \hspace{-0.2em} \cdots, \hspace{-0.2em} (K^n_v)_{\sigma_{k^n_v}}\}, \hspace{-0.1em} N, \hspace{-0.1em} t_{now})}$} 
		\myState {$\text{Verify}(\ac{LTC}_{ltca}, (tkt)_{\sigma_{ltca}})$}
		\vspace{0.1em}
		\myState {$tkt_{\sigma_{ltca}} \rightarrow (SN, H(Id_{PCA}\|Rnd_{tkt}),IK_{tkt}, t_s, t_e, Exp_{tkt})$}
		\vspace{0.1em}
		\myState {$H(Id_{this\textnormal{-}pca}\|Rnd_{n\textnormal{-}tkt}) \stackrel{?}{=} H(Id_{pca}\|Rnd_{n\textnormal{-}tkt})$}
		\vspace{0.1em}
		\myState {$Rnd_{v} \gets GenRnd()$}
		\For{i:=1 to \textbf{n}}{}
		\Begin
		\myState {$\text{Verify}(K^{i}_{v}, (K^i_v)_{\sigma_{k^i_v}})$} 
		\myState {${IK_{P^i_v} \gets H(IK_{tkt} || K^i_v || t_{s}^i || t_{e}^i|| H^{i}(Rnd_{v})})$}
		\If {$i = 1$} 
		\myState {$SN^i \gets H(IK_{P^i_v} || H^{i}(Rnd_{v}))$}
		\Else
		\myState {$SN^i \gets H(SN^{i-1} || H^{i}(Rnd_{v}))$}
		\EndIf
		\myState {${\zeta \leftarrow (SN^i, K^i_v, IK_{P^i_v}, t_{s}^i, t_{e}^i)}$} 
		\myState {$(P^i_v)_{\sigma_{pca}} \leftarrow Sign(Lk_{pca}, \zeta)$}
		\End
		\myState {\textbf{return} $(Id_{res}, \{(P^1_v)_{\sigma_{pca}}, \dots, (P^n_v)_{\sigma_{pca}}\}, Rnd_{v}, N\textnormal{+}1, t_{now})$}
		\EndProcedure
	\end{algorithmic}
\end{algorithm}

Having received a request, the \ac{PCA} verifies the ticket signed by the \ac{H-LTCA} (assuming trust is established between the two) (steps~\ref{protocol:scaling-vpkiaas-issuing-psnyms}.2\textendash\ref{protocol:scaling-vpkiaas-issuing-psnyms}.3). 
The \ac{PCA} then decapsulates the ticket and verifies the pseudonym provider identity (step \ref{protocol:scaling-vpkiaas-issuing-psnyms}.4\textendash\ref{protocol:scaling-vpkiaas-issuing-psnyms}.5). Then, the \ac{PCA} generates a random number (step~\ref{protocol:scaling-vpkiaas-issuing-psnyms}.6) and initiates a proof-of-possession protocol to verify the ownership of the corresponding private keys by the vehicle (step \ref{protocol:scaling-vpkiaas-issuing-psnyms}.9). Then, it calculates the \emph{``identifiable key''}, $IK: H(IK_{tkt} || K^i_v || t_{s}^i || t_{e}^i || H^{i}(Rnd_{v}))$ (step~\ref{protocol:scaling-vpkiaas-issuing-psnyms}.10). This essentially prevents a compromised \ac{PCA} from mapping a different ticket during resolution process, or identifies a malicious \ac{PCA} if issued a pseudonym without a valid ticket received. The \ac{PCA} implicitly correlates a batch of pseudonyms belonging to each requester (steps~\ref{protocol:scaling-vpkiaas-issuing-psnyms}.11\textendash\ref{protocol:scaling-vpkiaas-issuing-psnyms}.15). This essentially enables efficient distribution of the \ac{CRL}~\cite{khodaei2018VehicleCentric}: the \ac{PCA} only needs to include one entry per batch of pseudonyms without compromising their unlinkability. Finally, the \ac{PCA} issues the pseudonyms by signing it using its private key (steps~\ref{protocol:scaling-vpkiaas-issuing-psnyms}.16\textendash\ref{protocol:scaling-vpkiaas-issuing-psnyms}.17) and delivers the response (step~\ref{protocol:scaling-vpkiaas-issuing-psnyms}.19).

\begin{algorithm}[!t]
	\setcounter{equation}{0}
	\setstretch{0.99}
	\floatname{algorithm}{Protocol}
	\caption{Pseudonym Issuance Validation Process} 
	\label{protocol:scaling-vpkiaas-plausibility-check-for-the-pseudonym-issuance}
	{\scriptsize
		\vspace{-0.5em}
		\begin{align}
				V_j &: P^{i}_{v} \leftarrow (SN^i, K^i_v, IK_{P^i_v}, t_{s}^i, t_{e}^i) \\
				V_j &: \zeta \leftarrow (P^{i}_{v}) \\
				V_j &: (\zeta)_{\sigma_{v}} \leftarrow Sign(P^{j}_{v}, \zeta) \\
				V_j \rightarrow{\ac{RA}} &: (Id_{req}, (\zeta)_{\sigma_{v}}, t_{now}) \\
				\ac{RA} &: \text{Verify}(P_{v}, (\zeta)_{\sigma_{v}}) \\
				\ac{RA} &: \zeta \leftarrow (P^{i}_{v}) \\ 
				\ac{RA} &: (\zeta)_{\sigma_{ra}} \leftarrow Sign({Lk_{ra}}, \zeta) \\ 
				\ac{RA}\rightarrow{\ac{PCA}} &: (Id_{req}, (\zeta)_{\sigma_{ra}}, \ac{LTC}_{ra}, N, t_{now}) \\
				\ac{PCA} &: \text{Verify}(LTC_{ra}, (\zeta)_{\sigma_{ra}}) \\
				\ac{PCA} &: (tkt, Rnd_{IK_{P^{i}_{v}}}) \leftarrow \text{Resolve}(P^{i}_{v}) \\
				\ac{PCA} &: \chi \leftarrow (SN_{P^{i}}, tkt_{\sigma_{ltca}}, Rnd_{IK_{P^{i}_{v}}}) \\
				\ac{PCA} &: (\chi)_{\sigma_{pca}} \leftarrow Sign({Lk_{pca}}, \chi) \\ 
				\ac{PCA} \rightarrow \ac{RA} &: (Id_{res}, (\chi)_{\sigma_{pca}}, N\textnormal{+}1, t_{now}) \\ 
				\ac{RA} &: \text{Verify}(LTC_{pca}, \chi) \\ 
				\ac{RA} &: \hspace{-0.25em} (SN_{P^{i}}, tkt_{\sigma_{ltca}}, Rnd_{IK_{P^{i}_{v}}}) \hspace{-0.25em} \leftarrow \hspace{-0.25em} \chi \\ 
				\ac{RA} &: \text{Verify}(LTC_{ltca}, tkt_{\sigma_{ltca}}) \\ 
				\ac{RA} &: \hspace{-0.25em} (H(Id_{PCA}\|Rnd_{tkt}),IK_{tkt}, t^i_s, t^i_e, Exp_{tkt}) \hspace{-0.25em} \leftarrow \hspace{-0.25em} tkt \\ 
				\ac{RA} &: {H(IK_{tkt} || K^i_v || t^{i}_{s} || t^{i}_{e} || Rnd_{IK_{P^{i}_{v}}}) \stackrel{?}{=} IK_{P_v^i} } 
		\end{align}
		\vspace{-1em}
	}
\end{algorithm}

\subsubsection{\textbf{Pseudonym Issuance Validation Process (Protocol~\ref{protocol:scaling-vpkiaas-plausibility-check-for-the-pseudonym-issuance})}}
\label{subsubsec:scaling-vpkiaas-pseudonym-issuance-validation-process}

Upon receiving a request for misbehavior identification, e.g., multiple suspicious traffic congestion alerts sent to a traffic monitoring system, an entity could send a request to the \ac{RA} to validate the pseudonym issuance process of a \emph{``suspicious''} pseudonym (step~\ref{protocol:scaling-vpkiaas-plausibility-check-for-the-pseudonym-issuance}.1\textendash~\ref{protocol:scaling-vpkiaas-plausibility-check-for-the-pseudonym-issuance}.4). The \ac{RA} validates the request and interacts with the corresponding \ac{PCA} that issued the pseudonym, to provide evidence for the pseudonym issuance procedure; in fact, this process ensures that an actual vehicle requested the pseudonym by providing a valid ticket, also guarantees the \ac{PCA} did not issue a pseudonym for an illegitimate vehicle (step~\ref{protocol:scaling-vpkiaas-plausibility-check-for-the-pseudonym-issuance}.5\textendash~\ref{protocol:scaling-vpkiaas-plausibility-check-for-the-pseudonym-issuance}.8).

Upon receiving the request, the \ac{PCA} validates the request, and provides the corresponding ticket and $Rnd_{IK_{P^{i}_{v}}}$, used to issue the pseudonym. The response is signed by the \ac{PCA} sent back to the \ac{RA} (step~\ref{protocol:scaling-vpkiaas-plausibility-check-for-the-pseudonym-issuance}.8\textendash~\ref{protocol:scaling-vpkiaas-plausibility-check-for-the-pseudonym-issuance}.13). Upon receiving the response, the \ac{RA} verifies it, facilitates validating the ticket using the public key of the \ac{LTCA}, and checks ${H(IK_{tkt} || K^i_v || t^{i}_{s} || t^{i}_{e} || Rnd_{IK_{P^{i}_{v}}}) \stackrel{?}{=} IK_{P_v^i} }$ (step~\ref{protocol:scaling-vpkiaas-plausibility-check-for-the-pseudonym-issuance}.14\textendash~\ref{protocol:scaling-vpkiaas-plausibility-check-for-the-pseudonym-issuance}.18). If the hash calculation results in the same hash values, it confirms that the pseudonym has been issued based on a valid ticket, i.e., properly issued by the \ac{LTCA}. Moreover, it ensures the \ac{PCA} could not have issued the pseudonym for a \emph{non-existing} vehicle. Note that upon performing pseudonym issuance validation process, the actual identity of a vehicle is not disclosed, i.e., user privacy is strongly protected. Further security and privacy analysis in Sec.~\ref{sec:scaling-vpkiaas-security-and-privacy-analysis}.

\subsection{Mitigating Sybil Attacks on the \ac{VPKIaaS}}
\label{subsec:scaling-vpkiaas-sybil-protection}

Multiple replicas of a micro-service interact with the same database to accomplish their operations, e.g., all replicas of \acp{LTCA} should interact with the same database to store information about tickets they issue. The same way, all replicas of \acp{PCA} interact with a single database to validate an authorization ticket and store information corresponding to issued pseudonyms. Micro-services could opt in to utilize their shared MySQL database either synchronously or asynchronously\footnote{A synchronous interaction with a database implies enforcing limits on accessing to a resource by locking it to ensure the consistency of all transactions. An asynchronous interaction, though, implies that requests are proceeded without waiting to complete a transaction; the execution will happen later via an asynchronous callback function.}. Asynchronous interaction of the micro-services and the shared database would result in efficient pseudonyms issuance. However, a malicious vehicle could repeatedly submit requests. If the micro-services do not synchronously validate tickets and pseudonym requests, one can obtain multiple sets of pseudonyms if the requests were delivered to different replicas. On the other hand, synchronous interaction of the micro-services and the shared database would prevent issuing multiple sets of pseudonym for a given requester, thus, eradicating the Sybil-based misbehavior. However, it would drastically diminish the performance of the system, notably timely on-demand issuance of pseudonyms. The performance of the relational database, e.g., MySQL, used in~\cite{khodaei2018Secmace}, can be highly degraded by synchronized interactions, e.g.,~\cite{cooper2010benchmarking}. Moreover, scaling out the Pods to handle a large volume of workload while relying on a single shared MySQL database becomes a \emph{single point of failure}, questions the practicality of such a scheme (to be highly-available and dynamically-scalable).

\begin{figure}[!t] 
	\vspace{-2.5em}
	\centering
	\includegraphics[trim=0cm 0cm 0cm 0cm, clip=true, totalheight=0.31\textheight,angle=0,keepaspectratio]{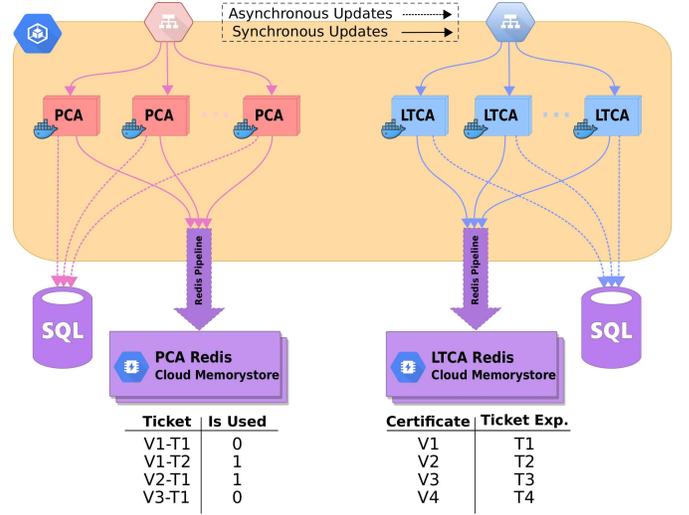}
	\vspace{-1.0em}
	\caption{\ac{VPKIaaS} Memorystore with Redis and MySQL.}
	\label{fig:scaling-vpkiaas-ltca-pca-redis-overview}
\end{figure}

In order to systematically mitigate the aforementioned vulnerability, we propose a hybrid design by considering two separate databases. Fig.~\ref{fig:scaling-vpkiaas-ltca-pca-redis-overview} shows the Memorystore of the \ac{VPKIaaS}: an in-memory key-value database as a service on \ac{GCP} compatible with the Redis~\cite{redis} protocol, and a relational database, e.g., MySQL. Each Pod of a micro-service synchronously interacts with the Redis database\footnote{Note that MySQL and Redis could both be single point of failures if not offered as a highly-available and dynamically-scalable service. However, a distributed cluster of MySQL will be a bottleneck in our scenario because relational databases are slow in nature, especially if the setup is synchronous. The Redis cluster, though, is an in-memory key-value database which offers very fast insertion and query.} to validate a request towards thwarting Sybil attacks. Upon validating a request, the tickets and pseudonyms are issued and the corresponding information are stored in the relational database asynchronously. Such a hybrid design mitigates Sybil attacks without diminishing the overall performance of the pseudonym acquisition process: the time-consuming validation through the rational database is replaced by an efficient validation through the Redis database.

\subsubsection{\ac{LTCA} Sybil Attack Mitigation (Protocol~\ref{protocol:scaling-vpkiaas-ticket-request-validation-by-redis})}
\label{subsubsec:scaling-vpkiaas-sybil-attack-mitigation-ltca}

The \ac{LTCA}, the \emph{policy decision and enforcement point} in a domain, issues tickets with non-overlapping intervals, i.e., vehicles cannot obtain tickets with overlapping lifetime. Upon receiving a ticket request, each \ac{LTCA} micro-service Pod should check if a ticket was issued to the requester during that period. Enforcing such a policy ensures that no vehicle would obtain more than a single valid ticket towards requesting multiple simultaneously valid pseudonyms. Furthermore, each ticket is implicitly bound to a specific \ac{PCA} by the vehicle; as a result, the ticket cannot be used more than once or be used for other \acp{PCA}. Each \ac{LTCA} micro-service Pod stores the serial number of the vehicle's \ac{LTC} (as the key) and the expiration time of its current ticket (as the value) on the Redis database. Upon receipt of a new request for obtaining a ticket, each micro-service creates a Redis pipeline to validate the ticket (step~\ref{protocol:scaling-vpkiaas-ticket-request-validation-by-redis}.2). A Redis pipeline entails a list of commands guaranteed to be executed sequentially without interruption.

\setlength{\textfloatsep}{0pt}
\begin{algorithm}[t!]
	\floatname{algorithm}{Protocol}
	\caption{Ticket Request Validation (by the \ac{LTCA} using Redis)}
	\label{protocol:scaling-vpkiaas-ticket-request-validation-by-redis}
	\algloop{For}{}
	\algblock{Begin}{End}
	\begin{algorithmic}[1]
		\Procedure{ValidateTicketReq}{$SN_{LTC}^{i}, tkt_{start}^{i}, tkt_{exp}^{i}$}
		\myState {$(value^i) \leftarrow \textnormal{RedisQuery} (SN_{LTC}^i)$}
		\If {$value^{i} == NULL \:\: \textnormal{\textbf{OR}} \:\: value^{i} <= tkt_{start}^{i}$} 
			\myState {$\textnormal{RedisUpdate} (SN_{LTC}^{i}, tkt_{exp}^{i})$} 
			\myState{$Status \leftarrow IssueTicket(\dots)$} \Comment{Invoking ticket issuance procedure}
			\If {$Status == False$} 
				\myState {$\textnormal{RedisUpdate} (SN_{LTC}^{i}, value^{i})$} \Comment{Reverting $SN_{LTC}^{i}$ to $value^{i}$}
				\myState {\textbf{return} $(False)$} \Comment{Ticket issuance failure}
			\Else
				\myState {\textbf{return} $(True)$} \Comment{Ticket issuance success}
			\EndIf
		\Else
			\myState {\textbf{return} $(False)$} \Comment{Suspicious to Sybil attacks}
		\EndIf
		\EndProcedure
	\end{algorithmic}
\end{algorithm}

The Redis pipeline checks the existence of the serial number of an \ac{LTC} in the database; if it exists, it validates if the request interval overlaps with the previously recorded entry (step~\ref{protocol:scaling-vpkiaas-ticket-request-validation-by-redis}.3); the request is marked to be \emph{malicious} if the serial number exists in the database and the requested ticket start time ($tkt_{start}$) is less than the expiration time of the already existed ticket. Otherwise, the Redis pipeline updates the corresponding entry (or adds a new entry if not existed) with the new ticket expiration time (step~\ref{protocol:scaling-vpkiaas-ticket-request-validation-by-redis}.4). Then, the procedure for ticket issuance will be invoked (step~\ref{protocol:scaling-vpkiaas-ticket-request-validation-by-redis}.5, i.e., Protocol~\ref{protocol:scaling-vpkiaas-issuing-ticket-algorithm} in Appendix). In case of any failure during the ticket issuance, the ticket expiration value will be rolled back (steps~\ref{protocol:scaling-vpkiaas-ticket-request-validation-by-redis}.6\textendash\ref{protocol:scaling-vpkiaas-ticket-request-validation-by-redis}.8). The Redis pipeline is executed on a single thread and it is guaranteed to sequentially execute the commands; thus, even if all replicas of the \ac{LTCA} received a ticket request from the same vehicle, Redis ensures that only one ticket request will be served and the rest of them will be denied.

\setlength{\textfloatsep}{0pt}
\begin{algorithm}[t!]
	\floatname{algorithm}{Protocol}
	\caption{Pseudonym Request Validation$\:$(by the \ac{PCA} using Redis)}
	\label{protocol:scaling-vpkiaas-pseudonym-request-validation-by-redis}
	\algloop{For}{}
	\algblock{Begin}{End}
	\begin{algorithmic}[1]
		\Procedure{ValidatePseudonymReq}{$SN_{tkt}^{i}$}
		\myState {$(value^i) \leftarrow \textnormal{RedisQuery} (SN_{tkt}^i)$}
		\If {$value^{i} == NULL \:\: \textnormal{\textbf{OR}} \:\: value^{i} == False$} 
			\myState {$\textnormal{RedisUpdate} (SN_{tkt}^{i}, True)$} 
			\myState{$Status \leftarrow IssuePsnyms(\dots)$} \Comment{Invoking pseudonym issuance}
			\If {$Status == False$} 
				\myState {$\textnormal{RedisUpdate} (SN_{tkt}^{i}, False)$} \Comment{Reverting $SN_{tkt}^{i}$ to False}
				\myState {\textbf{return} $(False)$} \Comment{Pseudonym issuance failure}
			\Else
				\myState {\textbf{return} $(True)$} \Comment{Pseudonym issuance success}
			\EndIf
		\Else
			\myState {\textbf{return} $(False)$} \Comment{Suspicious to Sybil attacks}
		\EndIf
		\EndProcedure
	\end{algorithmic}
\end{algorithm}

\subsubsection{\ac{PCA} Sybil Attack Mitigation (Protocol~\ref{protocol:scaling-vpkiaas-pseudonym-request-validation-by-redis})}
\label{subsubsec:scaling-vpkiaas-sybil-attack-mitigation-pca}

The \ac{PCA} issues pseudonyms with non-overlapping lifetimes in order to ensure that no vehicle is provided with more than one valid pseudonym at any given point in time. However, in order to fully eradicate Sybil-based misbehavior, the \ac{PCA} micro-service should ensure that each ticket is used only once to issue a set of pseudonyms for a requester. In other words, the \ac{VPKIaaS} system should ensure that different replicas of the \ac{PCA} micro-service never issue more than a set of pseudonyms for a ticket. All replicas of the \ac{PCA} share a Redis Memorystore with the ticket serial number (as the key) and a boolean data type (as the value). If the ticket serial number does not exist, or if it exists with a boolean data type value of false, the ticket was not used.

Upon receipt of a pseudonym acquisition request, each Pod of the \ac{PCA} micro-service creates a Redis pipeline to validate the ticket (step~\ref{protocol:scaling-vpkiaas-pseudonym-request-validation-by-redis}.2). If the key ($SN_{tkt}$) does not exist or the value is false (step~\ref{protocol:scaling-vpkiaas-pseudonym-request-validation-by-redis}.3), Redis updates the database with the value of true and the procedure for issuing pseudonyms will be invoked (step~\ref{protocol:scaling-vpkiaas-pseudonym-request-validation-by-redis}.5, i.e., Protocol~\ref{protocol:scaling-vpkiaas-issuing-psnyms}). In case of failure during the pseudonym acquisition process, the corresponding value for the ticket will be set to false in the Redis database, i.e., rolling back the transaction, to ensure the consistency of the pseudonym issuance procedure (steps~\ref{protocol:scaling-vpkiaas-pseudonym-request-validation-by-redis}.6\textendash\ref{protocol:scaling-vpkiaas-pseudonym-request-validation-by-redis}.8). If the value corresponding to the key ($SN_{tkt}$) is true, the request for obtaining a set of pseudonyms should be denied (step~\ref{protocol:scaling-vpkiaas-pseudonym-request-validation-by-redis}.13).


%% file: qualitative-analysis.tex
\section{Qualitative Analysis}
\label{sec:scaling-vpkiaas-security-and-privacy-analysis}

A detailed security and privacy analysis on the requirements for \ac{VPKI} entities can be found in~\cite{khodaei2018Secmace, khodaei2018VehicleCentric}. Here, we compile security and privacy analysis for deploying a \ac{VPKIaaS} system on the cloud, and we discuss additional facts of the problem. A detailed description on secret management in the cloud can be found in Appendix.

\subsection{Security and Privacy Analysis}
\label{subsec:security-privacy-analysis}

\emph{Sybil-based misbehavior:} A malicious vehicle could attempt to repeatedly request to obtain multiple tickets from the \ac{LTCA}, and/or aggressively request multiple sets of pseudonyms from the \ac{PCA}. However, all replicas of a micro-service share a Redis Memorystore to validate every request. Thus, any suspicious request can be \emph{instantaneously} validated through the Redis Memorystore (without interacting with the MySQL, which would be relatively more time-consuming). Redis is executed on a single thread and the pipeline is guaranteed to sequentially execute the commands; thus, even if all replicas of a micro-service, e.g., the \ac{PCA}, received a pseudonym request from one vehicle at the same time, the \ac{VPKIaaS} system would serve only one pseudonym request and the rest of them would be denied. Therefore, the \ac{VPKIaaS} ensures an efficient ticket and pseudonym provisioning while preventing any vehicle from obtaining multiple tickets or sets of pseudonyms towards a Sybil-based misbehavior. The ramification of the Redis service failure depends on the action taken after the failure, i.e., \emph{fail open} or \emph{fail close}. In case of fail open, Sybil attacks would be possible, as the \ac{VPKIaaS} system would provide vehicles with spurious pseudonyms. Later, it invalidates the erroneously issued credentials by adding them to the \ac{CRL}. In case of fail close, the \ac{VPKIaaS} system stops issuing credentials until the failure gets resolved.

Alternatively, a single deviant \ac{PCA} could issue multiple simultaneously valid pseudonyms for a given vehicle, or issue pseudonyms for an entity without any valid ticket issued by the \ac{LTCA}. However, upon performing pseudonym validation process, the \ac{RA} requests the corresponding \ac{PCA} to validate a pseudonym. Each pseudonym requires to have a valid pseudonym identifiable key ($IK_{P_v^i}$). Thus, a malicious \ac{PCA} can be identified and would then be evicted from the \ac{VPKI} system if it issued a pseudonym without a valid ticket provided. Note that when performing the pseudonym issuance validation process, the actual identity of the pseudonym owner is not disclosed to the \ac{PCA} or the \ac{RA}, i.e., user privacy is preserved. Moreover, no entity can infer user sensitive information by continuously conducting pseudonym issuance validation process towards harming user privacy. We emphasize here that our \ac{VPKIaaS} scheme does not prevent a malicious \ac{PCA} from issuing multiple sets of fake pseudonyms; rather, our scheme facilitates efficient identification of a misbehaving \ac{PCA} by cross-checking the pseudonym issuance procedure in a privacy-preserving manner. To ensure correct operation of a micro-service, each Pod frequently requests a dummy ticket or pseudonym. Since such operations are executed in isolation within the Pod, the issued dummy tickets and pseudonyms cannot leave the Pod. Moreover, each issued pseudonym can be cross-checked towards identifying suspicious compromised entity.

\emph{\ac{DDoS} attacks on the \ac{VPKIaaS} system:} Compromised internal entities or external adversaries could try to harm the system operations by launching a \ac{DoS} (or a \ac{DDoS}) attack, thus degrading the availability of the system. A rate limiting mechanism prevents them from compromising the availability of the system; moreover, the system flags misbehaving users, thus evicting them from the system. External adversaries could launch a \ac{DDoS} attack by clogging the \ac{LTCA} with fake certificates, or the \ac{PCA} with bogus tickets. In fact, such misbehaving entities attempt to compromise the availability of the \ac{VPKI} entities by mandating them to excessively validate the signature of fake \acp{LTC} or bogus tickets, i.e., performing a signature flooding attack~\cite{hsiao2011flooding}.

We achieve high-availability and fault-tolerance in the face of a benign failure by exploiting the Kubernetes master to kill the running (faulty) Pod, e.g., in case of system faults or crashes, and create a new Pod. In case of resource depletion attacks, the Kubernetes master scales out the Pods to handle such demanding loads. At the same time, a puzzle technique, e.g.,~\cite{aura2001resistant, abliz2009guided}, can be employed as a mitigation approach, e.g.,~\cite{khodaei2018Secmace}: each vehicle is mandated to visit a pre-defined set of Pods, in a pre-determined sequential order to solve a puzzle. As a result, the power of an attacker is degraded to the power of a legitimate client, thus, an adversary cannot send high-rate spurious requests to the \ac{VPKI}. On the side of the infrastructure, there are \ac{DDoS} mitigation techniques at different network layers, provided by various cloud service providers.

\emph{Synchronization among the \ac{VPKI} entities:} Lack of synchronization between the \ac{LTCA} and the \ac{PCA} could affect the pseudonym issuance process, e.g., a \ac{PCA} would not issue pseudonyms for a seemingly `expired' ticket. However, mildly drifting clocks of the \ac{VPKI} entities can hardly affect the operation, because the pseudonym lifetimes and periods for pseudonym refills ($\Gamma$) are in the order of minutes, typically. It suffices to have \ac{VPKI} entities periodically synchronizing their clocks. For example, if the accuracy of an \ac{RTC} is 50 parts-per-million (ppm), i.e., $50 \times 10^{-6}$, and the maximum accepted error in timestamp is 50 ms, then each entity should synchronize its clock every 16 minutes ($\frac{50 \times 10^{-3} sec}{50 \times 10^{-6} ppm}$).


%% file: quantitative-analysis.tex
\section{Quantitative Analysis}
\label{sec:scaling-vpkiaas-performance-evaluation}

\textbf{Experimental setup:} We leveraged a state-of-the-art \ac{VPKI} system~\cite{khodaei2018Secmace} and restructured its source code to fit in a micro-services architecture, e.g., through containerization, automation, bootstrapping of services. We further added health and load metric publishing features, to be used by an orchestration service to scale in/out accordingly. We built and pushed Docker images for \ac{LTCA}, \ac{PCA}, \ac{RA}, MySQL, and Locust~\cite{locust}, \emph{an open source load testing tool}, to the Google Container Registry~\cite{google-container-registry}. Isolated namespaces and deployment configuration files are defined before \ac{GKE} v1.10.11~\cite{google-kubernetes-engine} cluster runs the workload. We configured a cluster of five \acp{VM} (n1-highcpu-32), each with 32 vCPUs and 28.8GB of memory. The implementation is in C++ and we use FastCGI~\cite{heinlein1998fastcgi} to interface Apache web-server. We use XML-RPC~\cite{xmlrpc-c} to execute a remote procedure call on the cloud. The \acs{VPKIaaS} interface is language-neutral and platform-neutral, as we use Protocol Buffers~\cite{protocol-buffer} for serializing and de-serializing structured data. For the cryptographic protocols and primitives (\ac{ECDSA} and \acs{TLS}), we use OpenSSL with \ac{ECDSA}-256 key pairs according to the ETSI (TR-102-638)~\cite{ETSI-102-638} and IEEE 1609.2~\cite{1609-2016} standards; other algorithms and key sizes are compatible in our implementation.

To facilitate the deployment of the \ac{VPKIaaS}, we created all \ac{VPKIaaS} configuration in YAML language~\cite{yaml}, applicable to deploy on any cloud provider which offers \emph{Kubernetes As A Service}, e.g., \ac{GCP}~\cite{gcp} and \ac{AWS} (aws.amazon.com). For our experiments, we deployed our \ac{VPKIaaS} on the \ac{GKE}. We also used other \ac{GCP} services: \emph{Memorystore}~\cite{memorystore}, \emph{Prometheus}~\cite{prometheus}, and \emph{Grafana}~\cite{grafana}. The Memorystore service is a Redis-compatible~\cite{redis} service which acts as in-memory key-value data store (see Fig.~\ref{fig:scaling-vpkiaas-ltca-pca-redis-overview}). Prometheus is a feature-rich metric service which collects all the metrics of the Kubernetes cluster and the applications running on it into a time-series database. We use Grafana to visualize the metrics collected by Prometheus and monitor the \emph{system under test}. Prometheus and Grafana are deployed as prepared applications from the \ac{GCP} marketplace~\cite{prometheus-grafana} on the Kubernetes cluster. Moreover, we leveraged Locust~\cite{locust}, deployed on the Kubernetes cluster, to synthetically generate a large volume of pseudonym requests.

\begin{table}
	\vspace{-2em}
	\centering
	\caption{Experiment Parameters.}
	\vspace{-0.75em}
	\label{table:scaling-vpkiaas-experiments-parameters}
	\hspace{-0.5em}
	\resizebox{0.48\textwidth}{!}
	{
		\renewcommand{\arraystretch}{1.2}
		\begin{tabular}{ | c | c | c | }
			\hline
			\rowcolor{gray!80}
			\textbf{Parameters} & \textbf{Config-1} & \textbf{Config-2} \\\hline\hline
			\textbf{total number of vehicles} & 1000 & 100, 50,000 \\\hline 
			\rowcolor{gray!20} \textbf{hatch rate} & 1 & 1, 100 \\\hline
			\textbf{interval between requests} & 1000-5000 ms & 1000-5000 ms \\\hline 
			\rowcolor{gray!20} \textbf{pseudonyms per request} & 100, 200, 300, 400, 500 & 100, 200, 500 \\\hline
			\textbf{\ac{LTCA} memory request} & 128 MiB & 128 MiB \\\hline
			\rowcolor{gray!20} \textbf{\ac{LTCA} memory limit} & 256 MiB & 256 MiB \\\hline 
			\textbf{\ac{LTCA} CPU request} & 500 m & 500 m \\\hline 
			\rowcolor{gray!20} \textbf{\ac{LTCA} CPU limit} & 1000 m & 1000 m \\\hline 
			\textbf{\ac{LTCA} \acs{HPA}} & 1-40; CPU 60\% & 1-40; CPU 60\% \\\hline 
			\rowcolor{gray!20} \textbf{\ac{PCA} memory request} & 128 MiB & 128 MiB \\\hline 
			\textbf{\ac{PCA} memory limit} & 256 MiB & 256 MiB \\\hline 
			\rowcolor{gray!20} \textbf{\ac{PCA} CPU request} & 700 m & 700 m \\\hline 
			\textbf{\ac{PCA} CPU limit} & 1000 m & 1000 m \\\hline 
			\rowcolor{gray!20} \textbf{\ac{PCA} \acs{HPA}} & 1-120; CPU 60\% & 1-120; CPU 60\% \\\hline 
		\end{tabular}
	}
	\vspace{1em}
\end{table}

\textbf{Metrics:} To evaluate the performance of our \ac{VPKIaaS} system, we measure the latency to obtain pseudonyms under different scenarios and configurations for a large-scale mobile environment. More specifically, we evaluate the performance of the system with (and without) flash crowds to illustrate its \emph{high-availability}, \emph{robustness}, \emph{reliability}, and \emph{dynamic-scalability}. We demonstrate the performance of our \acs{VPKIaaS} system by emulating a large volume of synthetic workload. Table~\ref{table:scaling-vpkiaas-experiments-parameters} shows the configurations used in our experiments, with \emph{Config-1} referring to a \emph{`normal'} vehicle arrival rate and \emph{Config-2} for a \emph{flash crowd} scenario. Experiments with \emph{Config-1} indicates that every 1-5 seconds, a new vehicle joins the system and requests a batch of 100-500 pseudonyms. To emulate a flash crowd scenario, i.e., \emph{Config-2}, beyond having vehicles joining the system based on \emph{Config-1}, 100 new vehicles join the system every 1-5 seconds and request a batch of 100-200 pseudonyms. 

\textbf{Remark:} Assuming the pseudonyms are issued with non-over-lapping intervals (important to mitigate Sybil-based misbehavior), obtaining 100 and 500 pseudonyms per day implies pseudonyms lifetimes of 14.4 minutes ($\tau_{P}=14.4$ min.) or 3 minutes ($\tau_{P}=$172.8 sec), respectively. 
According to actual large-scale urban vehicular mobility dataset, e.g., Tapas-Cologne~\cite{uppoor2014generation} or \acs{LuST}~\cite{codeca2015lust}, the average trip duration is within 10-30 minutes. Moreover, according to the US \ac{DoT}, the average daily commute time in the US is around 1 hour~\cite{DOTHS812014}. Thus, requesting 100 pseudonyms per day would cover 24 hours trip duration with each pseudonym lifetime of approx. 15 minutes. We evaluate the performance of our \ac{VPKIaaS} system under such seemingly extreme configurations.

\subsection{Large-scale Pseudonym Acquisition}
\label{subsec:scaling-vpkiaas-large-scale-pseudonym-acquisition}

\begin{figure} [!t] 
	\vspace{-2em}
	\begin{center}
		\centering
		\subfloat[Ticket Issuance]{
			\hspace{-1.15em} \includegraphics[trim=0cm 0cm 0.5cm 0.95cm, clip=true, totalheight=0.15\textheight,angle=0,keepaspectratio]{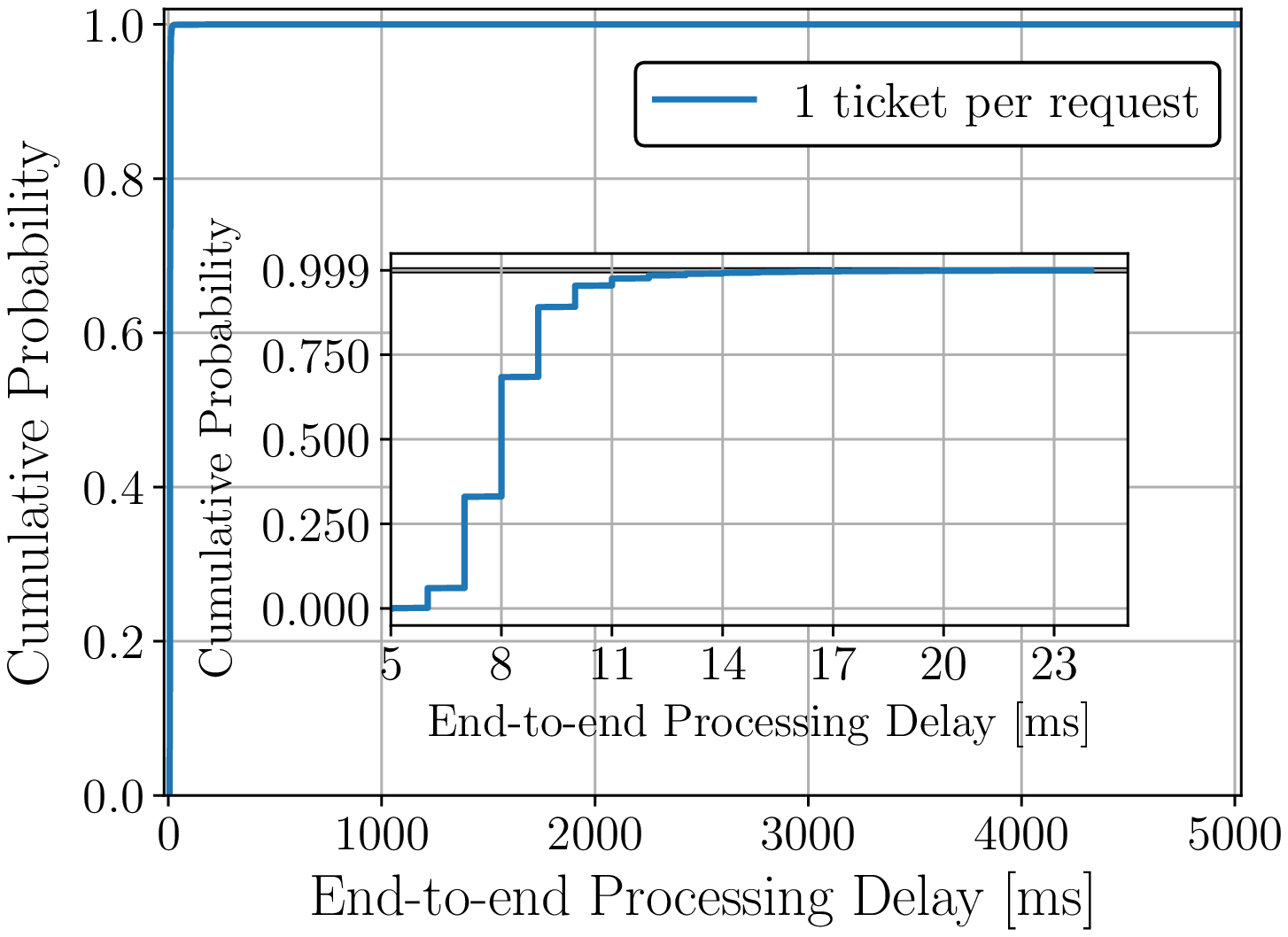}}
		\subfloat[Pseudonyms Issuance] {
			\hspace{-0.95em} \includegraphics[trim=0cm 0cm 0.75cm 0.95cm, clip=true, totalheight=0.15\textheight,angle=0,keepaspectratio]{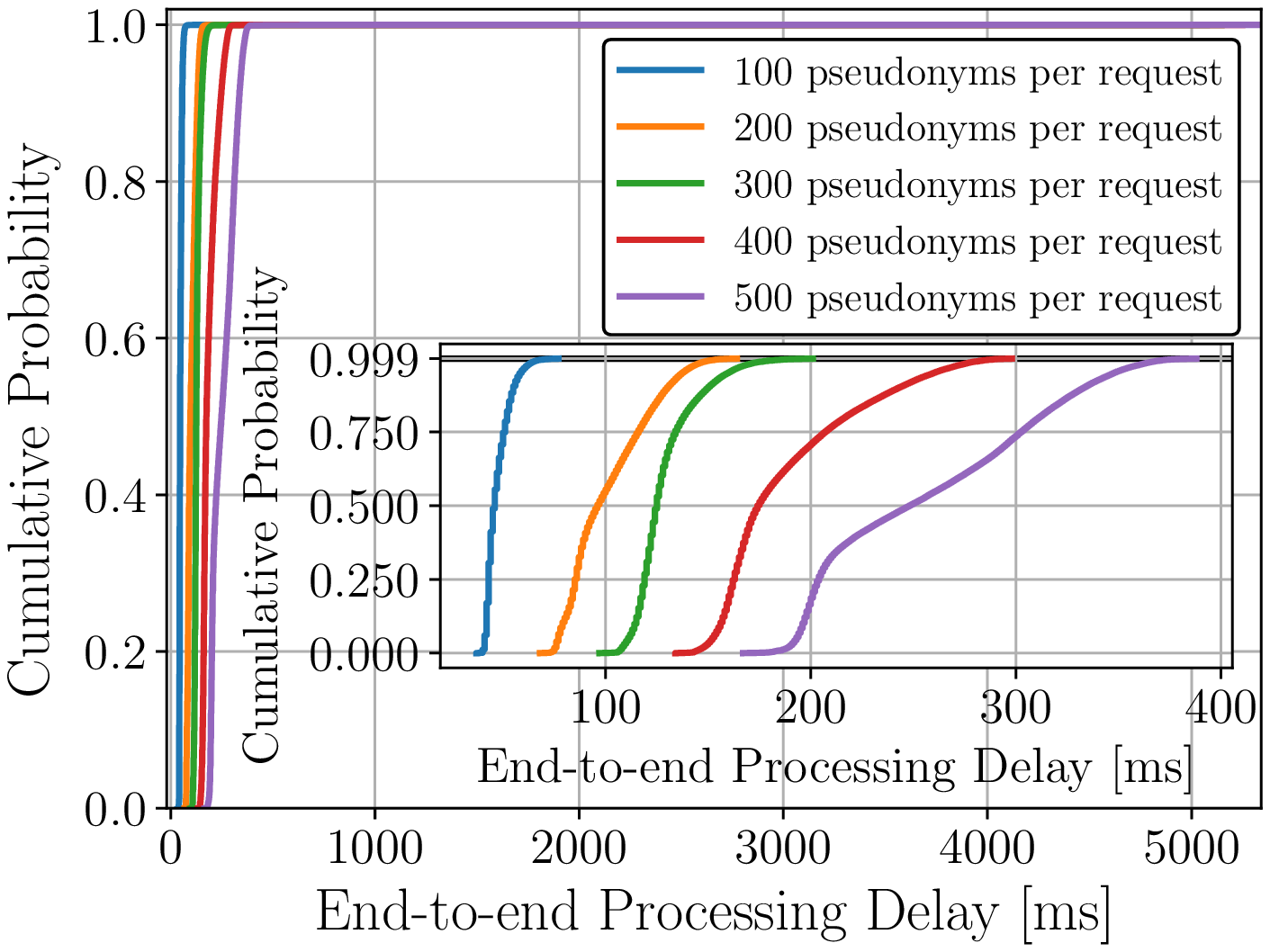}}
		\vspace{-0.75em}
		\caption{(a) CDF of end-to-end latency to issue a ticket. \\ (b) CDF of end-to-end processing delay to issue pseudonyms.}
		\label{fig:scaling-vpkiaas-end-to-end-latency-to-obtain-ticket-pseudonyms}
	\end{center}
	\vspace{-0.75em}
\end{figure}

Fig.~\ref{fig:scaling-vpkiaas-end-to-end-latency-to-obtain-ticket-pseudonyms}.a illustrates the \ac{CDF} of the single ticket issuance processing delay (executed based on Config-1 in Table~\ref{table:scaling-vpkiaas-experiments-parameters}); as illustrated, 99.9\% of ticket requests are served within 24 ms: $F_x(t=24 \: ms)=0.999$, i.e., $Pr\{t\leq24 \: ms\}=0.999$. Fig.~\ref{fig:scaling-vpkiaas-end-to-end-latency-to-obtain-ticket-pseudonyms}.b shows the \ac{CDF} of processing latency for issuing pseudonyms with different batches of pseudonyms per request as a parameter. For example, with a batch of 100 pseudonyms per request, 99.9\% of the vehicles are served within less than 77 ms ($F_x(t=77 \: ms)=0.999$). Even with a batch of 500 pseudonyms per request, the \ac{VPKIaaS} system can efficiently issue pseudonyms: $F_x(t=388 \: ms)=0.999$. The results confirm that the \ac{VPKIaaS} scheme is efficient and scalable: the pseudonym acquisition process incurs low latency and it efficiently issues pseudonyms for the requesters. 

\begin{figure} [!t] 
	\vspace{-1em}
	\begin{center}
		\centering
		\subfloat[]{
			\hspace{-0.75em} \includegraphics[trim=0cm 0cm 0.75cm 1.25cm, clip=true, totalheight=0.142\textheight,angle=0,keepaspectratio]{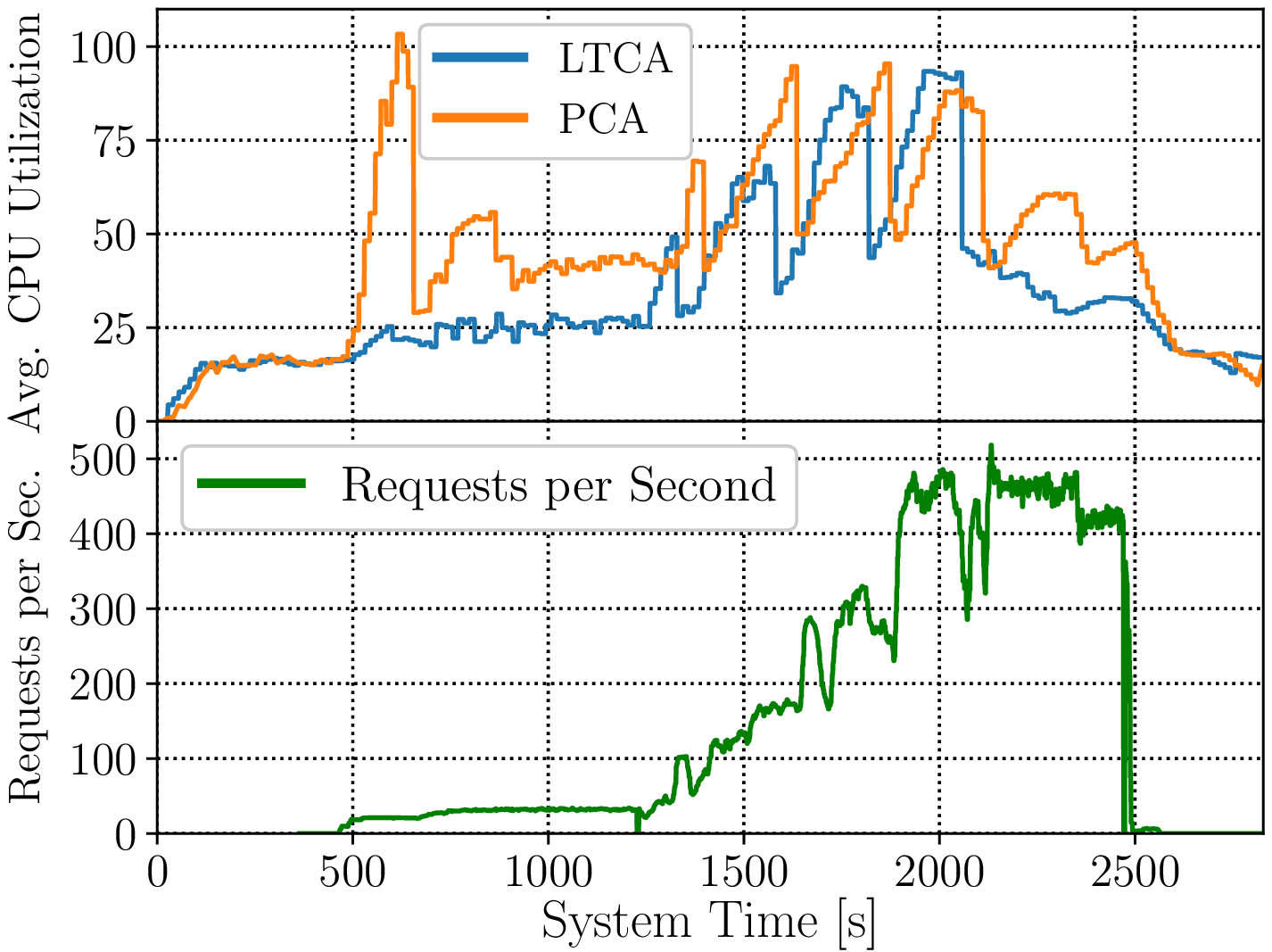}}
		\subfloat[] {
			\hspace{-0.75em} \includegraphics[trim=0cm 0cm 0.75cm 1.25cm, clip=true, totalheight=0.142\textheight,angle=0,keepaspectratio]{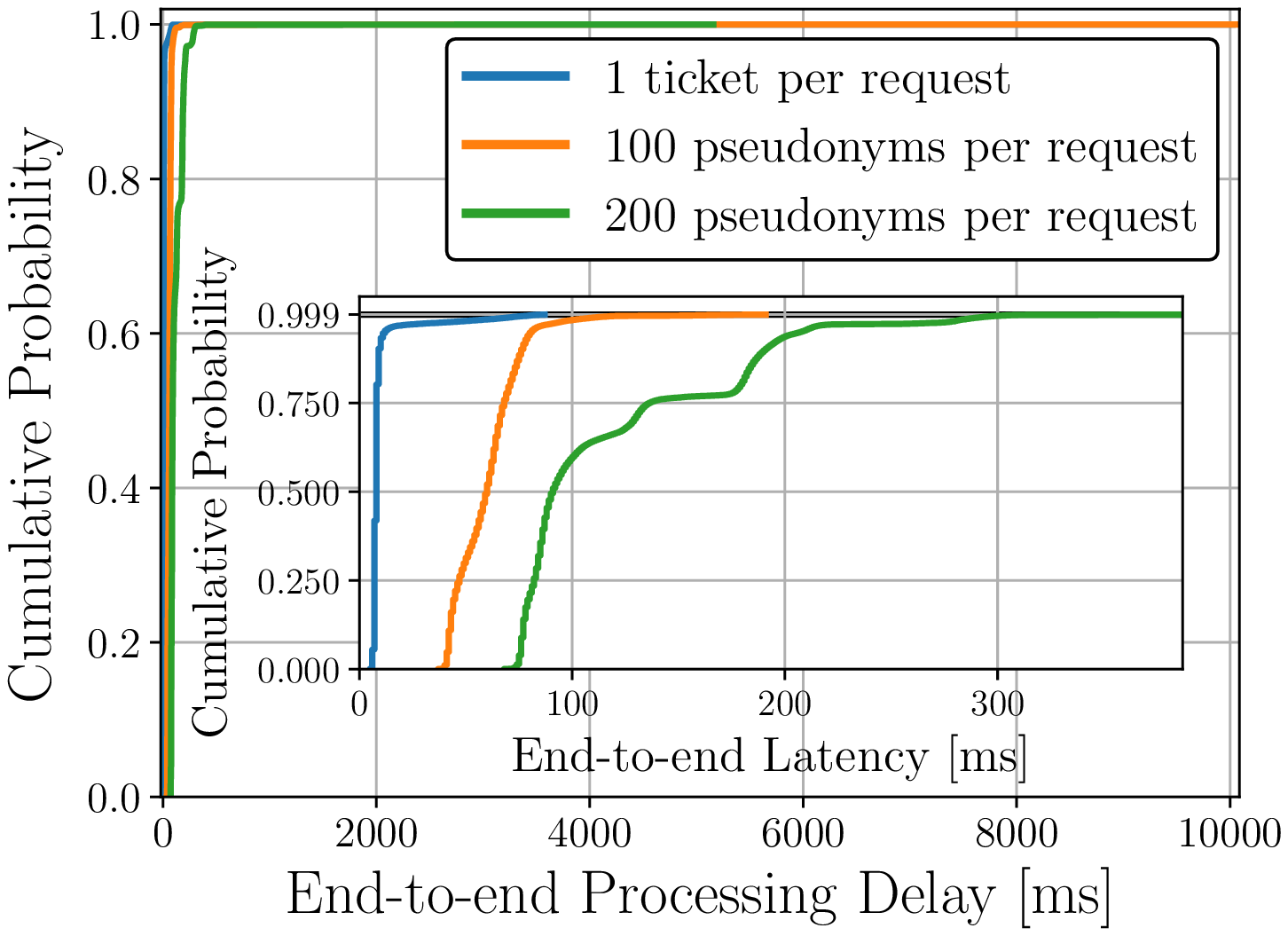}}
		\vspace{-0.75em}
		\caption{\ac{VPKIaaS} system in a flash crowd load situation. (a) CPU utilization and the number of requests per second. (b) CDF of processing latency to issue tickets and pseudonyms.}
		\label{fig:scaling-vpkiaas-high-availability-flash-crowd-load-pattern}
	\end{center}
	\vspace{0.25em}
\end{figure}

\subsection{\ac{VPKIaaS} with Flash Crowd Load Pattern}
\label{subsec:scaling-vpkiaas-high-availability-with-flash-crowd-load-pattern}

Fig.~\ref{fig:scaling-vpkiaas-high-availability-flash-crowd-load-pattern} shows the performance of the \ac{VPKIaaS} when a surge in pseudonym acquisition requests happens to the \ac{VPKIaaS} (executed based on Config-2 in Table~\ref{table:scaling-vpkiaas-experiments-parameters}, with 100 pseudonyms per request for Fig.~\ref{fig:scaling-vpkiaas-high-availability-flash-crowd-load-pattern}.a). We assess CPU utilization of the \ac{LTCA} and the \ac{PCA} Pods (Fig.~\ref{fig:scaling-vpkiaas-high-availability-flash-crowd-load-pattern}.a top) and the total number of pseudonyms requests per second (Fig.~\ref{fig:scaling-vpkiaas-high-availability-flash-crowd-load-pattern}.a bottom). When the number of requests per second increases, the average CPU utilization would rise; however, when CPU utilization hits 60\% threshold, defined in the \acp{HPA}~\cite{hpa}, the \ac{LTCA} and the \ac{PCA} deployment would horizontally scale to handle demanding loads, thus the average CPU utilization drops upon scaling out.

Fig.~\ref{fig:scaling-vpkiaas-high-availability-flash-crowd-load-pattern}.b shows the end-to-end processing latency to obtain tickets and a batch of 100 or 200 pseudonyms in a flash crowd situation. The processing latency to issue a single ticket is: $F_x(t=87 \: ms)=0.999$; to issue a batch of 100 pseudonyms per request, the processing latency is: $F_x(t=192 \: ms)=0.999$. In comparison with processing delay under `normal' conditions (Fig.~\ref{fig:scaling-vpkiaas-end-to-end-latency-to-obtain-ticket-pseudonyms}), the processing latency of issuing a single ticket increases from 24 ms to 87 ms; the processing latency to issue a batch of 100 pseudonyms increased from 77 ms to 192 ms. Thus, even under such a highly demanding request rate, the \ac{VPKIaaS} system issues credentials efficiently.\footnote{The total number of vehicles requesting 100 pseudonyms (under Config-2 in Table~\ref{table:scaling-vpkiaas-experiments-parameters}) is 398,870 and the \ac{VPKIaaS} system issued approximately 40 millions pseudonyms within 2,500 seconds; with such an arrival rate, the \ac{VPKIaaS} system would issue $0.5 \times 10^{12}$ pseudonyms per year. Obviously, this number is lower that the one mentioned in Sec.~\ref{sec:scaling-vpkiaas-introduction}, i.e., $1.5 \times 10^{12}$. Note that this is a proof of concept of the implementation and evaluation of the \ac{VPKIaaS} system; by allocating more resources and increasing the pseudonym request rates, the \ac{VPKIaaS} system would issue even further pseudonyms.}

\begin{figure} [!t] 
	\vspace{-2.25em}
	\begin{center}
		\centering
		\subfloat[]{
			\hspace{-1.25em} \includegraphics[trim=0cm 0cm 0.75cm 1.25cm, clip=true, totalheight=0.148\textheight,angle=0,keepaspectratio]{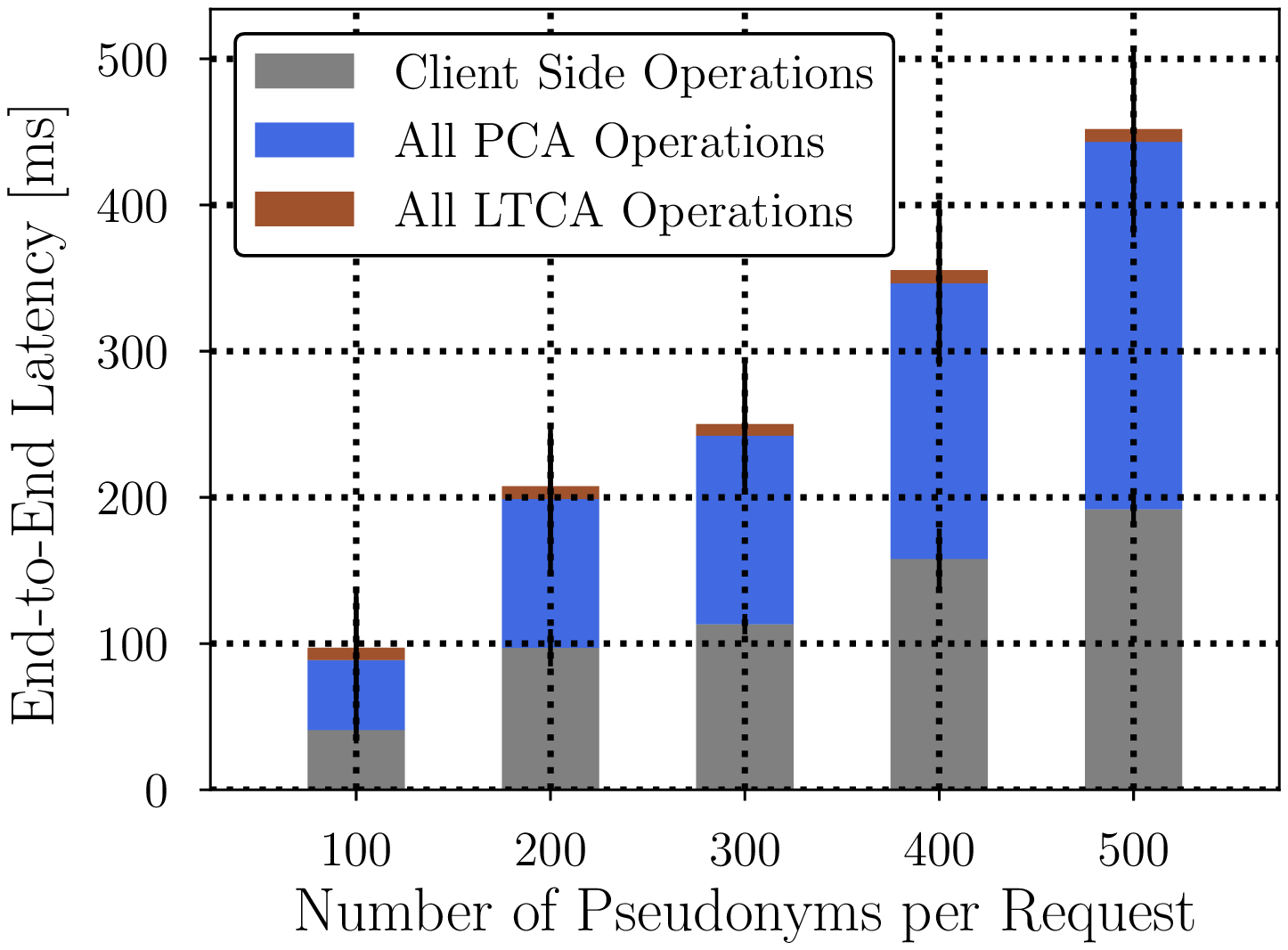}}
		\subfloat[] {
			\hspace{-1.25em} \includegraphics[trim=0cm 0cm 0.75cm 1.25cm, clip=true, totalheight=0.148\textheight,angle=0,keepaspectratio]{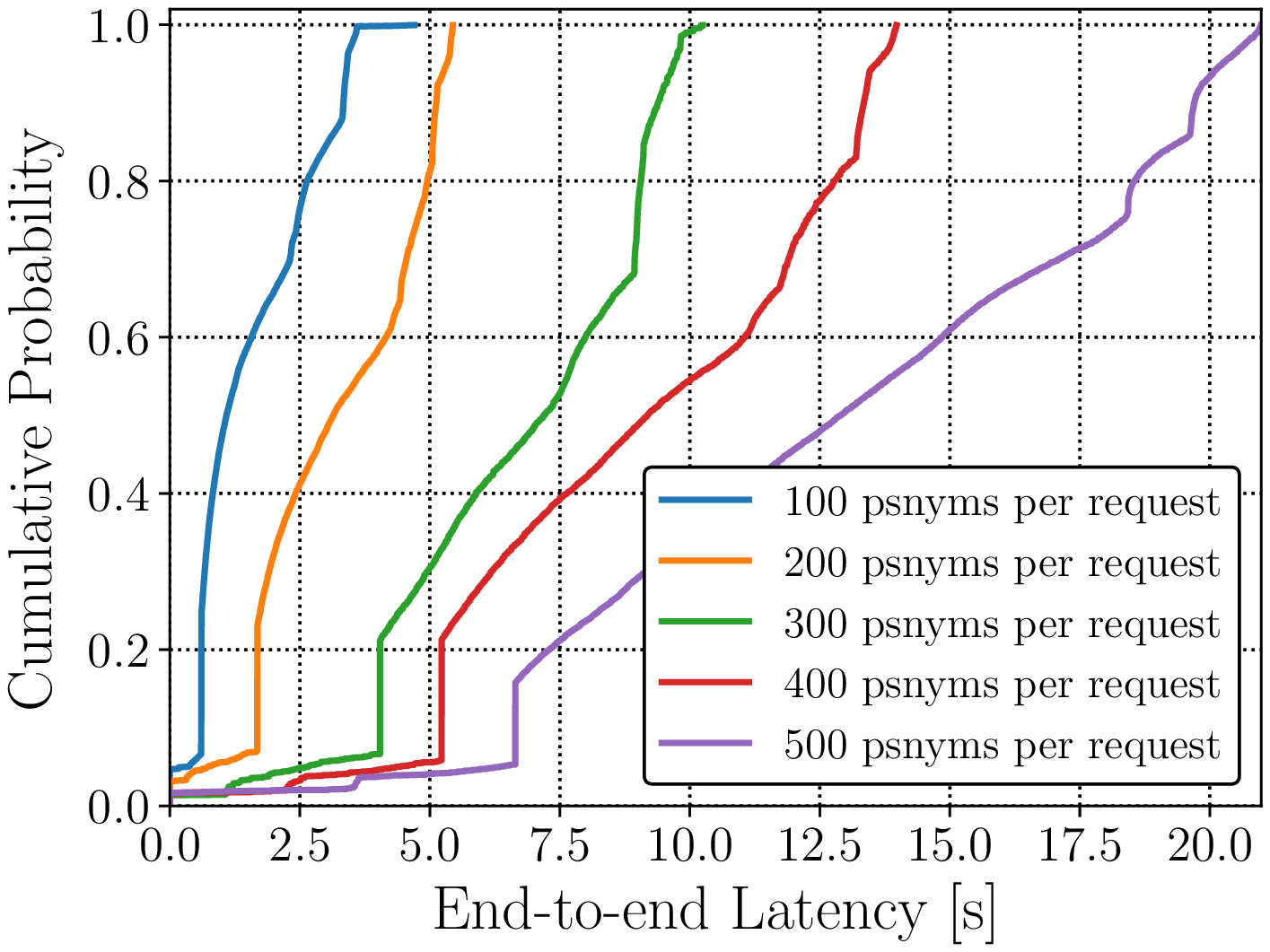}}
		\vspace{-0.75em}
		\caption{\ac{VPKIaaS} system with flash crowd load pattern. (a) Average end-to-end latency to obtain pseudonyms. (b) \ac{CDF} of end-to-end latency, observed by clients.}
		\label{fig:scaling-vpkiaas-high-availability-different-cpu}
	\end{center}
	\vspace{0em}
\end{figure}

Fig.~\ref{fig:scaling-vpkiaas-high-availability-different-cpu}.a shows the latency for each system component to obtain different batches of pseudonyms per request (Config-2 in Table~\ref{table:scaling-vpkiaas-experiments-parameters}). Our \ac{VPKIaaS} system outperforms prior work~\cite{cincilla2016vehicular}: the processing delay to issue 100 pseudonym for~\cite{cincilla2016vehicular} is approx. 2010 ms, while it is approx. 56 ms in our system, i.e., achieving a 36-fold improvement over prior work~\cite{cincilla2016vehicular}. Fig.~\ref{fig:scaling-vpkiaas-high-availability-different-cpu}.b illustrates the average end-to-end latency to obtain pseudonyms, observed by clients. As we can see, during a surge of requests, \emph{all} vehicles obtained a batch of 100 pseudonyms within less than 4,900 ms (including the networking latency). Obviously, the shorter the pseudonym lifetime, the higher the workload on the \ac{VPKI}, thus the higher the end-to-end latency. Note that serving requests under a flash crowd scenario at this rate (Config-2 in Table~\ref{table:scaling-vpkiaas-experiments-parameters}) implies that our \ac{VPKIaaS} system would serve 720,000 vehicles joining the system within an hour. Thus, even under such flash crowd load pattern, our \ac{VPKIaaS} system can comfortably handle such a high demand of requests.

\subsection{Dynamic-scalability of the \ac{VPKIaaS}}
\label{subsec:scaling-vpkiaas-dynamic-scalability-vpkiaas}

\begin{figure} [!t] 
	\vspace{-3em}
	\begin{center}
		\centering
		\subfloat[]{
			\hspace{-0.95em} \includegraphics[trim=0cm 0cm 0.75cm 0.35cm, clip=true, totalheight=0.154\textheight,angle=0,keepaspectratio]{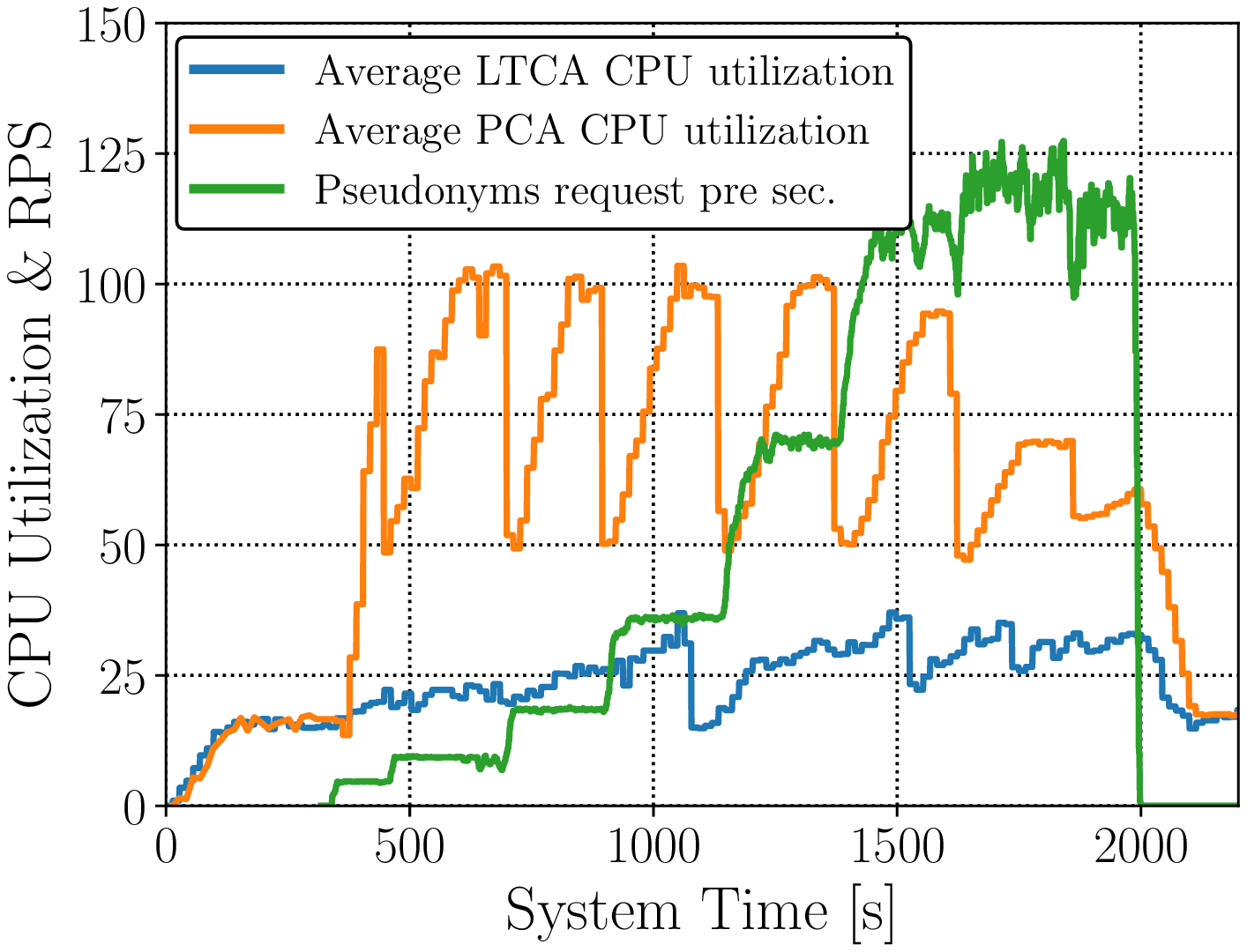}}
		\subfloat[] {
			\hspace{-0.85em} \includegraphics[trim=0.5cm 0.25cm 0.75cm 0.35cm, clip=true, totalheight=0.154\textheight,angle=0,keepaspectratio]{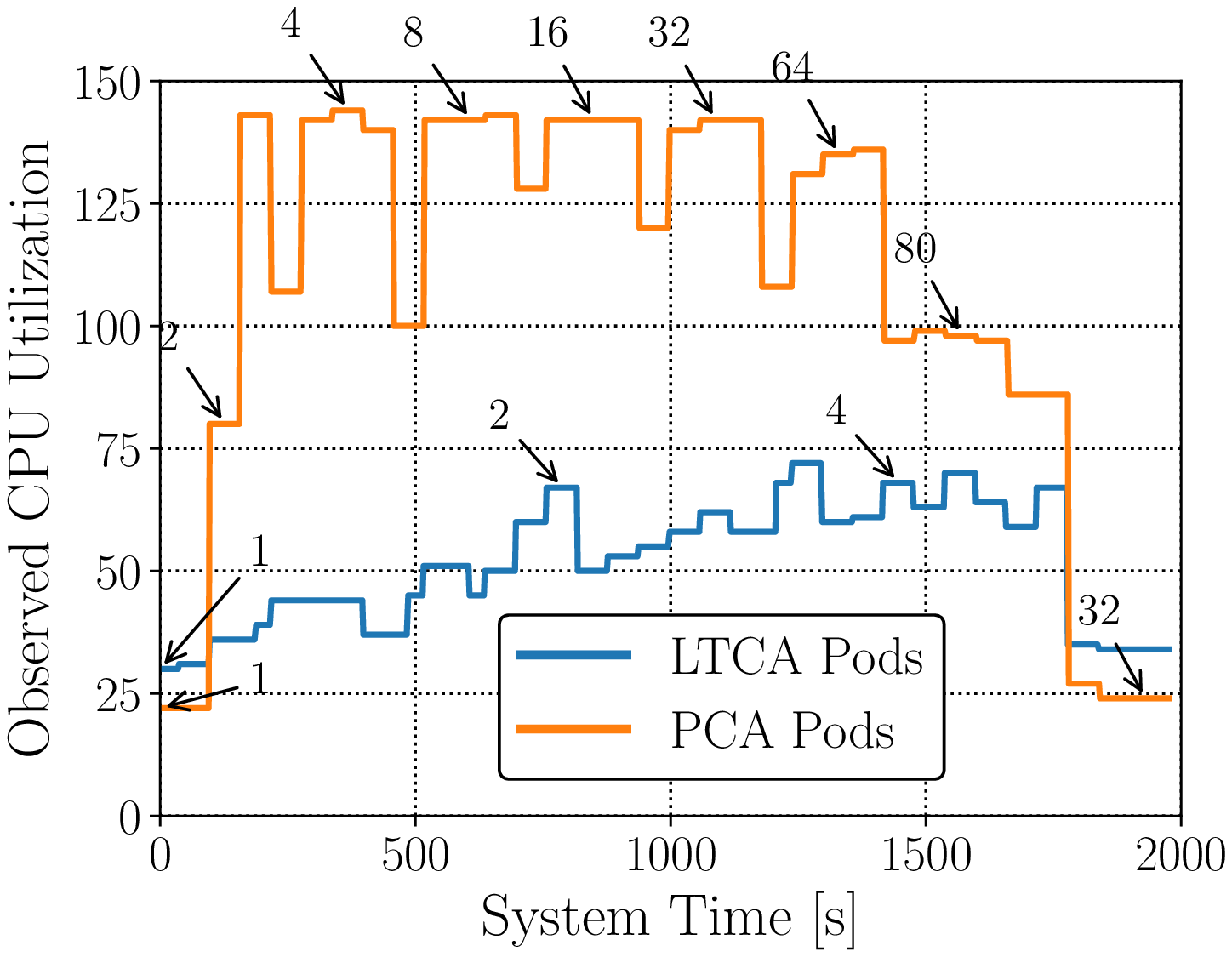}}
		\vspace{-0.75em}
		\caption{Each vehicle requests 500 pseudonyms (CPU utilization observed by \acs{HPA}). (a) Number of active vehicles and CPU utilization. (b) Dynamic scalability of \acs{VPKIaaS} system.} 
		\label{fig:scaling-vpkiaas-dynamic-scalability}
	\end{center}
	\vspace{0.5em}
\end{figure}

In this scenario, we demonstrate the performance of our \ac{VPKIaaS} system, notably its reliability and dynamic scalability. To emulate a large volume of workload, we generated synthetic workload using 30 containers, each with 1 vCPU and 1GB of memory (executed based on Config-2 in Table~\ref{table:scaling-vpkiaas-experiments-parameters}). Fig.~\ref{fig:scaling-vpkiaas-dynamic-scalability}.a shows the average CPU utilizations of the \ac{LTCA} and \ac{PCA} Pods (observed by \ac{HPA}) as well as the total number of requests per second. Fig.~\ref{fig:scaling-vpkiaas-dynamic-scalability}.b shows how our \ac{VPKIaaS} system dynamically scales out or scales in according to the rate of pseudonyms requests. The numbers next to the arrows show the number of \ac{LTCA} and \ac{PCA} Pod replicas at any specific system time. As illustrated, the number of \ac{PCA} Pods starts from 1 and it gradually increases; at system time 1500, there is a surge in pseudonym requests, thus the number of \ac{PCA} Pods increased to 80. Note that issuing a ticket is more efficient than issuing pseudonyms; thus, the \ac{LTCA} micro-service scaled out only up to 4 Pod replicas.

\subsection{\acs{VPKIaaS} Performance Comparison}
\label{subsec:scaling-vpkiaas-vpkiaas-performance-comparison}

We compare our \ac{VPKIaaS} scheme with a \emph{baseline} scheme~\cite{cincilla2016vehicular}, which implements a \ac{VPKI} according to the ETSI architecture. More precisely, each vehicle requests pseudonyms from an authorization authority; the request is forwarded to the enrollment authority to check and validate the request. Upon a successful validation, the authorization authority issues the pseudonyms and sends them back to the vehicle. Using the similar setup to have a meaningful and direct comparison, we achieve a 36-fold improvement over the baseline scheme: under normal conditions, the processing delay to issue 100 pseudonyms for the baseline scheme is approx. 2010 ms, while it is approx. 56 ms in our \ac{VPKIaaS} system. Even under a flash crowd scenario (based on Config-2), the processing delay to issue 100 pseudonyms is approx. 71 ms, i.e., 28-fold improvement. Furthermore, unlike the \ac{VPKI} system in~\cite{cincilla2016vehicular}, our implementation supports dynamic scalability, i.e., the \ac{VPKI} scales out, or scales is, based on the arrival rate of pseudonyms requests.

Moreover, in order to handle a large volume of workload, SECMACE~\cite{khodaei2018Secmace} requires to statically allocate resources to the \ac{VPKI}. In case of an unpredictable surge in the arrival rates or being under a \ac{DDoS} attack, the performance of SECMACE would drastically decrease. Furthermore, when deploying SECMACE on the cloud, a malicious vehicle could repeatedly request to obtain pseudonyms towards performing Sybil-based misbehavior. On the contrary, our \ac{VPKIaaS} system can comfortably handle requests with unexpected arrival rate while being \emph{efficient} in issuing pseudonyms, being \emph{resilient} against Sybil and resource depletion attacks, and being cost-effective by systematically allocating and deallocating resources.

%% file: conclusions.tex
\section{Conclusion}
\label{sec:scaling-vpkiaas-conclusion}

Paving the way for the deployment of a secure and privacy-preserving \ac{VC} system relies on deploying a special-purpose \ac{VPKI}. However, its success requires extensive experimental evaluation, to ensure viability (in terms of performance and cost). We leverage a state-of-the-art \ac{VPKI}, enhance its functionality, and migrate it into the \ac{GCP} to illustrate its availability, resiliency, and scalability towards a cost-effective \ac{VPKI} deployment. Through extensive security and privacy analysis, we show that the \ac{VPKIaaS} system fully eradicates Sybil-based misbehavior without compromising the efficiency of the pseudonym acquisition process. All these investigations would catalyze the deployment of the central building block of secure and privacy-preserving \ac{VC} systems.


%% file: acknowledgement.tex
\section*{Acknowledgement}
\label{sec:acknowledgement}

Work supported by the Swedish Foundation for Strategic Research (SSF) SURPRISE project and the KAW Academy Fellowship Trustworthy IoT project.


%% file: appendix.tex
\section*{Appendix}
\label{sec:appendix}


\setlength{\intextsep}{0.5pt}
\begin{algorithm}[!h]
	\floatname{algorithm}{Protocol}
	\caption{Ticket Request from the \acs{LTCA} (by the vehicle)}
	\label{protocol:scaling-vpkiaas-requesting-ticket-algorithm}
	\begin{algorithmic}[1]
		\Procedure{ReqTicket}{$t_{s}, t_{e}$}
		\myState {$Rnd_{tkt} \gets GenRnd()$}
		\myState {$\zeta \leftarrow (Id_{req}, H(Id_{PCA}\|Rnd_{tkt}), t_s, t_e)$}
		\myState {$(msg)_{\sigma_{v}} \leftarrow Sign(Lk_v, \zeta)$}
		\myState {\textbf{return} $((msg)_{\sigma_{v}}, \acs{LTC}_v, N, t_{now})$}
		\EndProcedure
	\end{algorithmic}
	\vspace{0.25em}
\end{algorithm}

\setlength{\intextsep}{0.5pt}
\begin{algorithm}[!h]
	\floatname{algorithm}{Protocol}
	\caption{Issuing a Ticket (by the \acs{LTCA})}
	\label{protocol:scaling-vpkiaas-issuing-ticket-algorithm}
	\begin{algorithmic}[1]
		\Procedure{IssueTicket}{$(msg)_{\sigma_{v}}, \acs{LTC}_v, N, t_{now}$}
		\myState {$\text{Verify}(\acs{LTC}_v, (msg)_{\sigma_{v}})$}
		\myState {$Rnd_{IK_{tkt}} \gets GenRnd()$}
		\myState {${IK_{tkt} \leftarrow H(\acs{LTC}_v || t_s || t_e || Rnd_{IK_{tkt}})}$}
		\myState {$\zeta \leftarrow (SN, H(Id_{PCA}\|Rnd_{tkt}),IK_{tkt}, t_s, t_e, Exp_{tkt})$}
		\myState {$(tkt)_{\sigma_{ltca}} \leftarrow Sign(Lk_{ltca}, \zeta)$}
		\myState {\textbf{return} $(Id_{res}, (tkt)_{\sigma_{ltca}}, Rnd_{IK_{tkt}}, N+1, t_{now})$}
		\EndProcedure
	\end{algorithmic}
\vspace{0.25em}
\end{algorithm}


\subsubsection*{\textbf{Ticket Acquisition Process (Protocols \ref{protocol:scaling-vpkiaas-requesting-ticket-algorithm} and \ref{protocol:scaling-vpkiaas-issuing-ticket-algorithm})}}
\label{subsubsec:scaling-vpkiaas-ticket-acquisition-process}

Assume the \ac{OBU} decides to obtain pseudonyms from a specific \ac{PCA}. It first interacts with its \ac{H-LTCA} to obtain a valid ticket. To conceal the actual identity of its desired \ac{PCA} from the \ac{LTCA}, it calculates the hash value of the concatenation of the specific \ac{PCA} identity with a random number\footnote{The storage cost for these random numbers is reasonably cheap, e.g., 264 million vehicles with average trip duration of 1 hour require 32 GB per day (25\$ per month).} (steps~\ref{protocol:scaling-vpkiaas-requesting-ticket-algorithm}.1\textendash\ref{protocol:scaling-vpkiaas-requesting-ticket-algorithm}.2). The vehicle prepares the request and signs it under the private key corresponding to its \ac{LTC} (step~\ref{protocol:scaling-vpkiaas-requesting-ticket-algorithm}.3\textendash\ref{protocol:scaling-vpkiaas-requesting-ticket-algorithm}.4) before returning the ticket request (step \ref{protocol:scaling-vpkiaas-requesting-ticket-algorithm}.5). It will then interact with the \ac{LTCA} over a bidirectional authenticated \ac{TLS}.

Upon reception of the ticket request, the \ac{LTCA} verifies the \ac{LTC} (thus authenticating and authorizing the requester) and the signed message (step \ref{protocol:scaling-vpkiaas-issuing-ticket-algorithm}.2). The \ac{LTCA} generates a random number ($Rnd_{IK_{tkt}}$) and calculates the \emph{``ticket identifiable key''} ($IK_{tkt}$) to bind the ticket to the \ac{LTC} as: $H(\acs{LTC}_v || t_s || t_e || Rnd_{IK_{tkt}})$ (steps \ref{protocol:scaling-vpkiaas-issuing-ticket-algorithm}.3\textendash\ref{protocol:scaling-vpkiaas-issuing-ticket-algorithm}.4); this prevents a compromised \ac{LTCA} from mapping a different \ac{LTC} during the resolution process. The \ac{LTCA} then encapsulates (step \ref{protocol:scaling-vpkiaas-issuing-ticket-algorithm}.5), signs (step \ref{protocol:scaling-vpkiaas-issuing-ticket-algorithm}.6), and delivers the response (step \ref{protocol:scaling-vpkiaas-issuing-ticket-algorithm}.7).

\subsection*{Secret Management}
\label{subsec:scaling-vpkiaas-secret-management}

Secret management is a concern towards deploying services in the cloud. Passwords, secret keys, and private keys cannot be simply integrated (hard-coded) into the services, e.g., the source code or the configuration files. Having services deployed on the cloud, each service fetches the needed secrets according to role-based access control. In this section, we review best practices, provided by cloud service providers. Note that deploying services on the cloud typically implies trusting cloud service providers, notably in terms of secret management.

\emph{\acf{AWS}:} \acs{AWS} offers several services regarding secret management on the cloud. The most common service is \ac{KMS}, which offers a key management service on \acs{FIPS} 140-2 validated \acp{HSM}~\cite{aws-cloudhsm} as the way to create, import, store, and rotate keys within \acs{AWS}. The \ac{KMS} of the \acs{AWS} only supports \ac{AES}-256. Through role-based access control policies for key management, one can be ensured that the secret key is only accessible by the authorized service, which initiated the process. As the applications and services will fetch the secrets from the \ac{KMS} whenever they need, changing the secret key will not affect the operations of the services because they will fetch a new secret in the next iteration. The \ac{KMS} provides automatic secret rotation, which can be enabled by the service. \ac{KMS} can also be integrated with CloudTrail~\cite{cloudtrail}, logging access to the secret keys. CloudTrail logs must be configured with proper actions, besides raising an alarm in case of suspicious activities, e.g., rotating the key if illegitimate access to the secret key. Beyond \ac{KMS}, there are other services for secret management, specifically designed to hold secret strings for the use in \ac{RDS} services of \acs{AWS}. \acs{AWS} also offers \ac{ACM}, providing a traditional certificate management~\cite{amazon-certificate-manager}.

\emph{\acf{GCP}:} \ac{GCP} offers a key management service similar to \acs{AWS}. Unlike \acs{AWS}, \ac{GCP} supports various cryptographic algorithms and primitives, e.g., \ac{AES}-256, RSA-2048, RSA-3072, RSA-4096, \ac{ECC}-P256, and \ac{ECC}-P384. In order to protect the secrets according to compliance standards, e.g., \ac{FIPS} 140-2, the \ac{KMS} can be integrated with the \ac{HSM}~\cite{google-cloudhsm} service provided by the cloud service provider according to the \ac{FIPS}~140-2 Level~3~\cite{carnahan1994security, fips-2018}. Accessing the secret keys can be restricted using \ac{IAM} policies~\cite{cloud-identity-access-management-iam}. The cloud \ac{IAM} service facilitates fine-grained access control to a service, e.g., defining a role to enable encryption using a certain \ac{KMS} service and assigning the role to a specific (\emph{authorized}) micro-service. Thus, the system ensures that only the specified micro-service can access the \ac{KMS} instance without being able to access other cryptographic materials. Similar to CloudTrail in \acs{AWS}, the \ac{GCP} provides an Audit Logging Service~\cite{google-cloud-audit-logging} in order to monitor activities, e.g., accessing the data, as well as logging the system events for auditing purposes.

\emph{Kubernetes:} Kubernetes is an orchestration service, responsible for orchestrating micro-services. Secret management in Kubernetes is different from the ones that cloud providers would offer. Kubernetes offers a secret management system for micro-services. Thus, a micro-service can leverage only the secret management system within the Kubernetes, or alternatively, it can interact with the secret management services offered by the cloud service provider, in which the Kubernetes instances operate.

\emph{Secret Management in VPKI\lowercase{aa}S:} In order to offer a cloud agnostic solution for the \ac{VPKIaaS} system, the Kubernetes secret volume suits our solution the best. However, the contents of the volume are encrypted using the \ac{KMS} of the cloud service provider. During the bootstrapping phase, each Pod of a micro-service fetches its encrypted private key from its local volume; it then queries the \ac{KMS} of the cloud provider to decrypt the private key (according to role-based access control). To protect the secrets, i.e., the key-pairs used by the \ac{VPKI} entities, each micro-service leverages its own secret volume in its own namespace~\cite{namespace}. A namespace is an isolated environment with all classes of elements operating in it. However, to protect the secret volumes, the keys can be encrypted using the \ac{KMS} of the cloud service provider, depending on the choice of deployment.
